\documentclass{aa}  

\usepackage{graphicx}
\usepackage{txfonts}
\usepackage{natbib}
\usepackage{xcolor}
\usepackage[hyperindex,breaklinks=true, colorlinks, citecolor=blue]{hyperref}  

\newcommand{\dd}{\mbox{d}}

\begin{document} 

   \title{The RWST, a comprehensive statistical description of the non-Gaussian structures in the ISM}
   \titlerunning{The RWST, a comprehensive statistical description of the non-Gaussian structures in the ISM}

\author{E. Allys\inst{\ref{inst1}}, F. Levrier\inst{\ref{inst1}}, S.~Zhang\inst{\ref{inst2},\ref{inst3}}, C. Colling\inst{\ref{inst5}},  B. Regaldo-Saint Blancard\inst{\ref{inst1}}, F. Boulanger\inst{\ref{inst1}}, P. Hennebelle\inst{\ref{inst5},\ref{inst1}}, S.~Mallat\inst{\ref{inst2},\ref{inst6}}}
\authorrunning{E. Allys et al.}

\institute{Laboratoire de Physique de l’Ecole normale supérieure, ENS, Université PSL, CNRS, Sorbonne Université, Université Paris-Diderot, Sorbonne Paris Cité, Paris, France\label{inst1} \and {Data team, \'Ecole Normale Sup\'erieure, Universit\'e PSL, 45 rue d'Ulm, 75005 Paris, France\label{inst2}} \and {centre for Data Science, Peking University, Beijing, China\label{inst3}} \and {Laboratoire AIM, IRFU/Service d'Astrophysique - CEA/DSM - CNRS - Universit\'e Paris Diderot, B\^at. 709, CEA-Saclay, F-91191 Gif-sur-Yvette Cedex, France\label{inst5}} \and {Collège de France, 11 place Marcelin Berthelot, 75005, Paris, France\label{inst6}}}

   \date{Received December 24th ;\\ Accepted July 7th.}

\abstract
{The interstellar medium (ISM) is a complex nonlinear system governed by the interplay between gravity and magneto-hydrodynamics, as well as radiative, thermodynamical, and chemical processes. Our understanding of it mostly progresses through observations and numerical simulations, and a quantitative comparison between these two approaches requires a generic and comprehensive statistical description of the emerging structures. The goal of this paper is to build such a description, with the purpose of permitting an efficient comparison that is independent of any specific prior or model.
We started from the Wavelet Scattering Transform (WST), a low-variance statistical description of non-Gaussian processes, which was developed in data science and encodes long-range interactions through a hierarchical multiscale approach based on the wavelet transform. We performed a reduction of the WST through a fit of its angular dependencies. This allowed us to gather most of the information it contains into a few components whose physical meanings are identified and describe for instance isotropic and anisotropic behaviours.
The result of this paper is the reduced wavelet scattering transform (RWST), a statistical description with a small number of coefficients that characterises complex structures arising from nonlinear phenomena, in particular interstellar magnetohydrodynamical (MHD) turbulence, independently of any specific priors. The RWST coefficients encode moments of order up to four, have reduced variances, and quantify the couplings between scales. To show the efficiency and generality of this description, we applied it successfully to the following three kinds of processes that are a priori very different: fractional Brownian motions, MHD simulations, and {\it Herschel} observations of the dust thermal continuum in a molecular cloud. With fewer than 100 RWST coefficients when probing six scales and eight angles on 256 by 256 maps, we were able to perform quantitative comparisons, infer relevant physical properties, and produce realistic synthetic fields.}


   \maketitle
   

\section{Introduction}

The interstellar medium (ISM) serves as a good example of how complex natural physical systems can be. Its physics involve a highly nonlinear interplay of gravity and magnetohydrodynamics (MHD), as well as radiative, thermodynamical, and chemical processes~\citep{draine_11}. This complexity precludes the advent of a comprehensive model of the ISM, whose understanding mostly progresses empirically through observations, numerical simulations, and phenomenological models. Those approaches each benefit from continually improving observational capabilities~\citep[see, e.g.][]{schinnerer2013,pabst2017,pety2017,cormier2018} and computational power~\citep[see, e.g.][]{gent2013,hennebelle-2018,hopkins2018}. A key point of ISM studies therefore lies in the quantitative comparison between observational and simulated data, which has to be done statistically. To perform such a comparison, however, requires to properly characterise non-Gaussian processes with long-range correlations that are a consequence of the complex nonlinear physics at play. This must also be done using statistical descriptions that keep a reasonably low dimensionality in order to be of sensible use~\citep{donoho2000high}.

In recent years, the complex physics of the ISM has also become closely related to observational cosmology since some cosmological signals of interest are much smaller than the emission from Galactic foregrounds. An example of this is the search for B-modes of polarisation in the cosmic microwave background (CMB). A potential detection of this signal would constrain inflation models in the very early Universe. However, it cannot be conclusive unless the submillimetre polarised thermal emission from Galactic dust is properly accounted for~\citep{pb2015}. This, in turn, requires a statistical model of this foreground emission in order to optimise component separation methods and reliably quantify uncertainties affecting the expected primordial signal~\citep{planck2016-l04}. Such models exist, but only as phenomenological ones~\citep{vansyngel-et-al-2017}, which are hampered by simplifying assumptions, for example that the random (turbulent) component of the Galactic magnetic field may be described as a Gaussian process. The development of a comprehensive statistical model of the ISM is therefore not only a goal for Galactic astrophysics. There are also important implications in cosmology.

It is often possible to find specific statistical estimators to test a given phenomenological model and estimate its parameters. In this class of estimators, one may think of diagnostics of the intermittent dissipation of turbulence in the probability distribution functions (PDFs) of velocity fluctuations at small scales~\citep{frisch1995turbulence, falgarone-et-al-2009}, of the evolutionary state of molecular clouds based on column-density PDFs~\citep{kainulainen-et-al-2009}, of the relative contributions of solenoidal and compressive modes of turbulence from spectro-imaging moment maps~\citep{orkisz-et-al-2017}, and of the relative orientation of interstellar filaments and magnetic fields with dedicated histograms~\citep{planck2015-XXXV} or more evolved tools~\citep{jow-et-al-2018}. These tailored statistical descriptions allow us to characterise some specific non-Gaussian features, but they are of limited scope when no prior model is available.

Other statistical descriptions are not specifically designed to test a phenomenological model, but rather to characterise the morphology of the observed fields\footnote{In this paper, we use the word \emph{field} to describe the two-dimensional physical quantities under study that our statistical descriptions are applied to. This unifies different words that could be used in other communities, such as 'image', 'texture', or 'flow'.}. In this category, we find descriptions in terms of filaments, sheets, and voids based on Morse theory~\citep{sousbie-2011}, hierarchical structure analyses using dendrograms~\citep{houlahan-scalo-1992,rosolowsky-et-al-2008}, or detections of linear structures, such as the Rolling Hough Transform~\citep{clark-et-al-2014}. The stability of these descriptions under small deformations of the fields is, however, not easy to ensure. 

An inherent difficulty to statistically modelling ISM processes in a comprehensive way lies in their long-range interaction properties. In this case, a description using probability distributions of pixel values must be based on conditional probabilities involving many points, and this is not easily tractable. A simpler way to describe such processes involves a hierarchical multiscale approach: the small scale interactions lead to the formation of local structures at intermediate scales, that in turn interact to form structures at larger scales, etc. This requires properly separating the variability of the process under study at different scales, which is precisely the purpose of the wavelet transform~\citep{cohen1995wavelets,van2004wavelets,farge2010multiscale,farge2015wavelet}. 

Second-order moments of wavelet coefficients are closely related to standard power-spectrum approaches~\citep{flandrin1992wavelet,meyer1999wavelets,farge2010multiscale}. As a first step, some physical insight into ISM processes may be gained from these power spectrum analyses. They discriminate between different models of turbulence that make various assumptions about the compressibility of the fluid and the presence of a magnetic field~\citep[see, e.g.][]{K41,iroshnikov-1964,kraichnan-1965a,kraichnan-1965b,sridhar-goldreich-1994,goldreich-sridhar-1995,kowal-lazarian-2007,falceta-goncalves-et-al-2014}. However, second-order moments do not fully describe the statistical properties of non-Gaussian fields. Higher-order statistical moments have also been used, such as bispectra~\citep{burkhart2009density} or structure functions~\citep{she-leveque-1994}, but these are prone to exhibit high variances due to outliers, and are therefore of limited use when only limited good quality data is available. 

To beat these shortcomings, recent advances in data science have shown that it is possible to extract non-Gaussian features of fields in the multiresolution framework provided by the wavelet transform, while keeping a reduced variance (see Sec.~\ref{PartWSTProp}). The wavelet scattering transform~\citep[WST,][]{mallat2012group}, which makes use of the properties of directional wavelets, is inspired by the architecture of convolutional neural networks, and yields state-of-the-art results for image classification problems, without requiring any training stage~\citep{bruna2013invariant,sifre2013rotation}. The outputs of the WST, called \emph{scattering coefficients}, constitute an efficient, low-variance, low-dimensionality statistical description of non-Gaussian processes. They contain information on moments of order higher than two, are able to capture long-range correlations, and can be related to physical properties of the systems under study.

We note that studies of ISM emission maps making a direct use of the wavelet transform, and therefore related to the work presented here, have also been conducted. For instance, \cite{khalil-et-al-2006} used the wavelet transform modulus maxima method~\citep{mallat-zhong-1992} to analyse {\sc Hi} maps from the Canadian Galactic Plane Survey~\citep{taylor-et-al-2003} in terms of their multifractal spectrum and local, scale-dependent anisotropies. More recently, \cite{robitaille-et-al-2014} used complex Morlet wavelets on thermal dust emission maps from the Hi-GAL {\it Herschel} survey~\citep{molinari-et-al-2010}, to separate their Gaussian and non-Gaussian components by thresholding on the probability distribution function (PDF) of the wavelet coefficients, finding in particular that the non-Gaussian part correlates well with the molecular gas emission. These approaches are in some sense akin to studying the first layer of the WST that we describe in Sect. 2.2.

The goal of this paper is to make use of this new method borrowed from data science to statistically characterise the complex structures of the ISM. With this purpose in mind, this paper introduces a statistical description of even lower dimension, the reduced wavelet scattering transform (RWST), that is obtained from the WST through the identification of the different angular modulations of the scattering coefficients, whose physical meanings are identified. This reduction does not require specific priors, but assumes that the angular dependency is smooth, as expected for physical systems. We show it to be successful in characterizing very different processes: fractional Brownian motions~\citep{stutzki1998fractal}, column density maps generated from MHD simulations, and an observation of the Polaris Flare molecular cloud with the {\it Herschel} satellite~\citep{miville2010herschel}. The RWST allows us to perform quantitative comparisons between these processes, and to produce realistically looking synthetic fields.

The paper is organised as follows: Section~\ref{PartGlobalWST} offers a simplified presentation of the WST aimed at a general audience of physicists. Section~\ref{PartReducedScatCoeff} introduces the RWST and discusses the generality of the angular reduction that is performed. It also presents a validation of this approach through the synthesis of random fields based on the WST and RWST coefficients. Section~\ref{PartVariousTermsDiscussion} reviews the various components of the RWST coefficients, and gives examples of what physical features are encoded in these coefficients. Our conclusions and some perspectives for future works are presented in Sec.~\ref{PartConclusion}. Five appendices complete the paper. The basic properties of Morlet wavelets are given in Appendix~\ref{AppendixMorletWavelets}; the three different classes of physical fields used to build examples are described in Appendix~\ref{AppendixFlowsStudied}; some comments on the generalizations and limits of the RWST are given in Appendix~\ref{AppendixDetailsDuFit}; the possibility to achieve a local statistical description of fields with the reduced scattering coefficients, as well as the difficulties it poses, are discussed in Appendix~\ref{AppendixLocalWST}; finally, additional examples of RWST for different processes are given in Appendix~\ref{AppendixCompleteSets}.

\section{Global wavelet scattering transform}
\label{PartGlobalWST}

The starting point of the statistical description introduced in this paper is the wavelet transform. Its ability to perform an efficient scale separation allows to study processes with long-range interactions by means of a hierarchical multiscale approach, and the progressive identification of structures at different scales that interact with each other~\citep{farge2010multiscale}. From this, the WST has been built specifically to quantify these couplings between such structures. 

\subsection{Introduction}
\label{PartIntroWST}

Since its first introduction in data science~\citep{mallat2012group}, the WST has led to state-of-the-art classification results for handwritten digits and texture discrimination~\citep{bruna2013invariant}, including the most difficult textures databases~\citep{sifre2013rotation}. It has also been applied to quantum chemical energy regression and the prediction of molecular properties~\citep{eickenberg2018solid}. The goal of this section is to present and synthesise for a general audience of physicists the construction and the properties of the WST coefficients\footnote{The scattering coefficients are computed with a MATLAB  software called \texttt{scatnet} that is publically available (\texttt{https://www.di.ens.fr/data/software/scatnet/}). We also developed a python code to perform the WST.}. The results of this section are thus not new in themselves, but are formulated as much as possible in the language of physics rather than applied mathematics. A more complete presentation and discussion of this transform can be found in~\cite{bruna2013invariant} and~\cite{bruna2015intermittent}.

The initial purpose of the WST was to understand and reproduce the image classification successes obtained by deep-learning architectures, but by means of a statistical description that does not require any training stage, and is entirely controlled. The WST is built by successive convolutions of the field with Morlet wavelets followed by the application of the modulus operator. We mainly use in this paper the WST to characterise the global statistical properties of a field. The WST thus computes a set of global coefficients that can be labelled with the scales they characterise. It is however also possible to use the WST to achieve a local description of a field that is not statistically homogeneous, as presented in Appendix~\ref{AppendixLocalWST}. 

\subsection{Computation of the WST coefficients}
\label{PartComputationWSTCoeff}

We consider here a real-valued two-dimensional field $I(\boldsymbol x)$. Typically, $I(\boldsymbol x)$ is defined on a grid of $d\times d$ pixels and represents, for instance, an intensity level at a given wavelength in an astrophysical observation. In that case, $\boldsymbol x$ stands for a position in the sky. All the sizes discussed henceforth refer to certain numbers of pixels, that can in turn be related to physical lengths. We use discretised wavelets, defining a number $J$ of scales to consider. The integer scales $j$ are labelled from 0 to $J-1$ and correspond to effective sizes of $2^{j}$ pixels (we work accordingly with base-2 logarithms in the whole paper). Therefore, $2^J$ must be smaller than or equal to the size $d$ of the image.

The angles~$\vartheta$ are also labelled by integers $\theta$, such that 
\begin{equation}
\label{EqDefTheta}
\vartheta = (\theta -1) \cdot \pi /\Theta,
\end{equation}
where $\Theta$ denotes the number of angles in which we divide a $\pi$ interval\footnote{Note that the integer labels $j$ and $\theta$ are sometimes abusively identified in this paper with the effective scale $2^j$ in pixels and angle $\vartheta$ in radians they correspond to.}. As we work with real fields, it is indeed sufficient to consider the WST coefficients for angles $\vartheta$ in $[0,\pi)$, i.e. with $\theta$ going from 1 to $\Theta$. The redundancy of the other angles stems from the fact that the Fourier transform of a real-valued field $I(\boldsymbol x)$ verifies $\tilde{I}(- \boldsymbol k) =\tilde{I}^*(\boldsymbol k)$ where ${}^*$ stands for complex conjugation\footnote{In the whole paper, $\tilde{f}(k)$ is the Fourier transform of $f(x)$.}. From now on, we work with a labelling in terms of oriented scales $(j,\theta)$, each of which corresponds to a certain wavevector $\boldsymbol k$. This labelling will be particularly useful to distinguish scale and angular dependencies of the scattering coefficients.

\begin{figure}[t]
\begin{center}
\includegraphics[width = 0.48\textwidth]{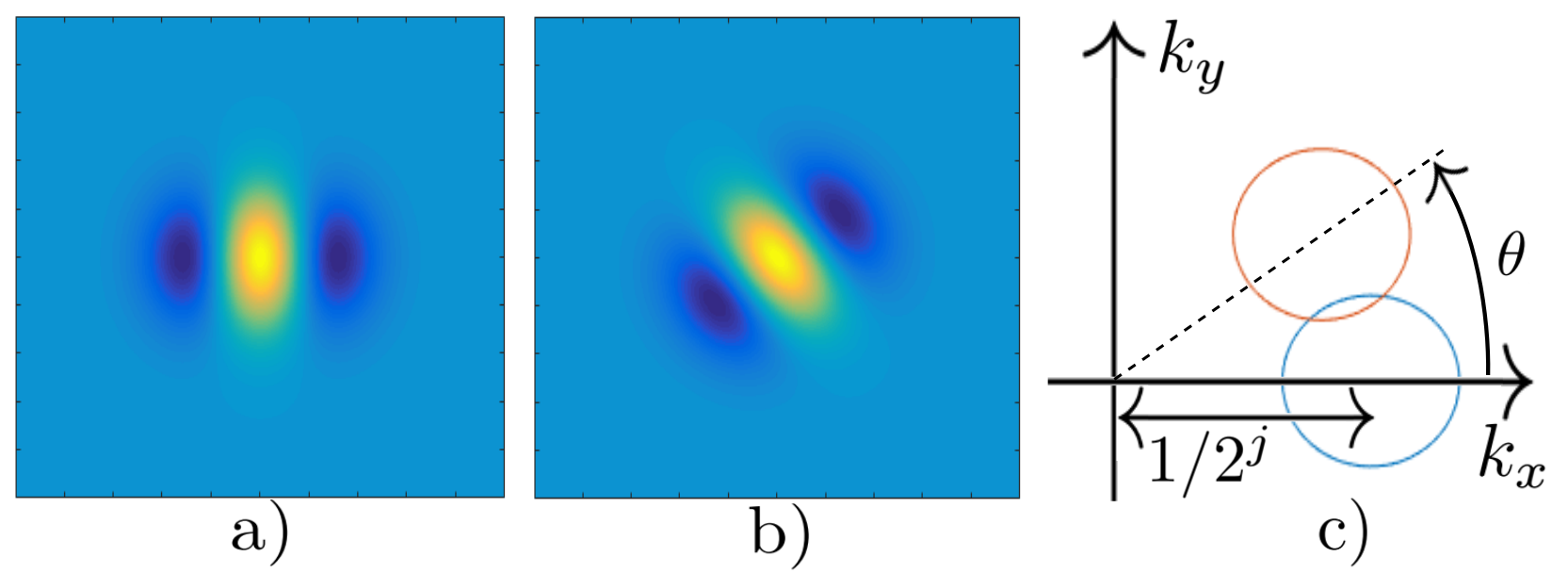}
\vspace{-0.5cm}
\end{center}
 \caption{Two-dimensional Morlet wavelets, with $\sigma=1$ (see Appendix~\ref{AppendixMorletWavelets}). (a)~Real part of $\psi_{j,0}$. (b)~Real part of $\psi_{j,\theta}$ with $\vartheta=\pi/5$. (c)~Location of the modulus of $\tilde{\psi}_{j,0}$ (blue) and $\tilde{\psi}_{j,\theta}$ (orange). We note that in our applications, we take $\sigma=0.8$, see Eq.~\eqref{EqDefMorlet1D}.}
\label{FigWavelet2D}
\end{figure}

The computation of scattering coefficients implemented in {\tt scatnet} involves convolutions with complex Morlet wavelets, which are convenient to interpret in the usual framework of spectral analysis since those wavelets are well localized in Fourier space. Their definition and basic properties are given in Appendix~\ref{AppendixMorletWavelets}, where their link with discrete windowed Fourier transforms is explained. Starting from an initial mother wavelet $\psi(\boldsymbol x)$ defined as the product of an oscillation of unit frequency and a Gaussian window [Eq.~\eqref{EqMotherWavelet2D}] the complex Morlet wavelets are then computed as
\begin{equation}
\label{EqWavelets2D}
\psi_{j,\theta}(\boldsymbol x) = 2^{-2j} \cdot \psi(2^{-j} r_{\theta}^{-1} \boldsymbol x),
\end{equation}
where $r_{\theta}$ is the rotation operator of angle $\theta$. The real parts of two examples of such wavelets and the supports of their Fourier transforms are shown in Fig.~\ref{FigWavelet2D}. The Fourier transform of the mother wavelet $\widetilde{\psi}(\boldsymbol{k})$ being centred on $\boldsymbol{k} _\psi$ with a unit bandwidth, $\tilde{\psi}_{j,\theta}$ has a support centred on $2^{-j}r_\theta \boldsymbol{k}_\psi$ with a bandwidth proportional to $2^{-j}$. For given values of $J$ and $\Theta$, one can build an appropriate set of wavelets $\{\psi_{j,\theta}\}$ such that their combined Fourier supports cover the whole Fourier plane, except for a localised area close to the null frequency\footnote{The lowest spatial frequencies can be probed by a dedicated Gaussian window.}~\citep{bruna2013invariant}. 

Using these wavelets, the WST coefficients are computed in three layers, indexed by an integer $m$ going from~0 to~2. The first layer $m=0$ characterises the average value of the field, and thus contains only one coefficient $S_0$
\begin{equation}
\label{EqCoeffm0}
S_0 =  \frac{1}{\mu_0} \int I(\boldsymbol{x}) ~ \dd^2 \boldsymbol{x},
\end{equation}
where the normalization factor $\mu_0$ is the surface over which the integration is performed. The coefficients $S_1(j_1,\theta_1)$ of the second layer $m=1$ depend on a single oriented scale $(j_1,\theta_1)$ and are given by
\begin{equation}
\label{EqCoeffm1}
S_1 (j_1,\theta_1) =  \frac{1}{\mu_1} \int \left| I \star \psi_{j_1,\theta_1} \right| (\boldsymbol{x}) ~ \dd^2 \boldsymbol{x},
\end{equation}
where $\star$ stands for the convolution and the normalization factor is the impulse response
\begin{equation}
\label{EqNormalizeMu1}
\mu_1 = \int \left| \delta \star \psi_{j_1,\theta_1} \right| (\boldsymbol{x}) ~ \dd^2 \boldsymbol{x},
\end{equation}
with $\delta$ the Dirac delta function\footnote{We do not write explicitly here the $j_i$ and $\theta_i$ dependencies of the $\mu_1$ and $\mu_2$ normalization factors.}. These $S_1$ coefficients probe the amplitudes of the spectral components of the field centred on the wavevector $2^{-j_1}r_{\theta_1} \boldsymbol{k}_\psi$ that is associated with the $(j_1,\theta_1)$ oriented scale. Finally, the coefficients $S_2 (j_1,\theta_1,j_2,\theta_2)$ of the third layer $m=2$ depend on two oriented scales $(j_1,\theta_1)$ and $(j_2,\theta_2)$, and are given by
\begin{equation} 
\label{EqCoeffm2}
S_2 (j_1,\theta_1,j_2,\theta_2) =    \frac{1}{\mu_2} \int \left| \left| I \star \psi_{j_1,\theta_1} \right| \star \psi_{j_2,\theta_2} \right| (\boldsymbol{x}) ~  \dd^2 \boldsymbol{x},
\end{equation}
where $ \mu_2$ is defined similarly to $\mu_1$ in Eq.~\eqref{EqNormalizeMu1}. These $S_2$ coefficients probe the level at which the first oriented scale $(j_1,\theta_1)$ is modulated at a second oriented scale $(j_2,\theta_2)$, with $j_2>j_1$ (see Sec.~\ref{PartWSTNumberNorm}). They are also related to geometrical shapes and structures appearing in the field~\citep{bruna2013invariant}. 
 
\subsection{Properties of the WST coefficients}
\label{PartWSTProp}

The WST coefficients depend on high-order moments of $I(\boldsymbol x)$, mainly of order up to $2^m$ for the $m^\text{th}$ layer~\citep{bruna2013invariant}. We therefore expect the $m=2$ coefficients to allow to distinguish fields that have the same second order moments (i.e. power spectra), but different higher order moments.

However, unlike high-order moments, whose estimators exhibit variances that are increasingly dominated by outliers, that is by samples which are far away from the mean~\citep{welling2005robust}, the WST coefficients do not involve products of values of the field. On the contrary, the WST coefficients are built using unitary and non-expansive operators (as the modulus), and have reduced variance, which means that they can be better estimated from limited a number of samples~\citep{bruna2013invariant}.

We note that the construction of scattering coefficients can be pursued for deeper $m\geqslant3$ layers, but in practice this is not necessary, and we choose to limit the present study to the $m\leqslant2$ layers. The $m\geqslant3$ layers describe couplings between three scales or more, and characterise accordingly correlations of order higher than four. It has however been shown in practice that those additional layers do not significantly improve the classification results or the quality of syntheses performed with the WST, despite an important increase in the number of scattering coefficients~\citep{bruna2013invariant}. 

Furthermore, for appropriate wavelets\footnote{By this we mean that the set of wavelet supports should cover the whole spectrum of the field under study, up to its largest scale. For example, for a $256\times256$ image, it requires to have $J=8$. Under this condition, Eq.~\eqref{EqEnergyConservation} is valid~\citep{mallat2012group}. In our case, we consider only the energy contained in the scales $j\leqslant 5$.}, the WST also preserves the field energy to a very good approximation~\citep{mallat2012group}
\begin{equation}
\label{EqEnergyConservation}
{|| I ||} ^2  = {S_0} ^2 +\sum_{j_1,\theta_1} {S_1}^2 (j_1,\theta_1)
+ \sum_{j_1,\theta_1,j_2,\theta_2} S_2^2 (j_1,\theta_1,j_2,\theta_2) + \varepsilon,
\end{equation}
where
\begin{equation}
\label{EqDefNormeL2}
{|| I ||} ^2 =  \frac{1}{{\mu_0}^{2}} \int | I(\boldsymbol x) |^2 \dd^2 \boldsymbol{x}.
\end{equation}
In Eq.~\eqref{EqEnergyConservation}, the $\varepsilon$ term stands for the energy encoded in the $m\geqslant 3$ layers, that contain in general less than $1\%$ of the initial energy of the field~\citep{bruna2013invariant}. Under the same requirement as for \eqref{EqEnergyConservation}, the conservation of energy can also be written at the level of the power spectrum:
\begin{equation}
\label{EqPSFromSC}
{|| I \star \psi_{j_1,\theta_1}||^2}  = {S_1}^2 (j_1,\theta_1)
+ \sum_{j_2,\theta_2} {S_2}^2 (j_1,\theta_1,j_2,\theta_2) + \varepsilon^\prime,
\end{equation}
where $|| I \star \psi_{j_1,\theta_1}||^2$ is defined following Eq.~\eqref{EqDefNormeL2}, and essentially represents the power spectrum of the field $I$ at the $(j_1,\theta_1)$ oriented scale [see Eq.~\eqref{EqDefPS} in Appendix~\ref{AppendixMorletWavelets}]. In this case, $\varepsilon^\prime$ also encodes the energy contained in the $m\geqslant3$ layers, that has been shown to be negligible for stationary processes~\citep{bruna2013invariant}. Eq.~\eqref{EqPSFromSC} shows that it is possible to recover the power spectrum of a field from its scattering coefficients. In addition, these properties link the distribution of energy into the different layers of the WST to the sparsity of the wavelet coefficients. Indeed, as $I \star \psi_{j_1,\theta_1}$ gets sparser, $S_1$ coefficients get smaller, and more energy is propagated to deeper layers. Highly non-Gaussian fields thus have larger $S_2$ coefficients, while the power spectrum of Gaussian fields may be recovered from the $S_1$ coefficients alone.

The WST finally has particular properties related to translations and small deformations of the field. First, the scattering coefficients are invariant under any global translation, since the coefficients are obtained after a spatial integration. Such a property is indeed required when working with homogenous statistics. Second, the WST linearises small deformations~\citep{mallat2012group}. This means that starting from a field $I(\boldsymbol x)$ and deforming it by a small amount\footnote{A proper formulation of this property requires the introduction of distances between two fields as well as between two sets of scattering coefficients. It is then possible to show that when deforming a field by a small amount, the displacement in the space of WST coefficients can be bounded in terms of the displacement in the field space. See~\cite{mallat2012group} or~\cite{bruna2013invariant} for more details.}, the associated displacement in the scattering coefficients space is bounded, and thus the modification of the statistical description performed by the WST is related in amplitude to the deformation of the field. Such a property is of prime importance when studying complex physical phenomena, since one expects two fields related by a small deformation to have similar physical properties. 

\subsection{Number and normalization of the WST coefficients}
\label{PartWSTNumberNorm}

The assumed values of $J$ and $\Theta$ determine the number of scattering coefficients describing a given field. Let us first note that the $S_2$ coefficients are negligible for $j_2 < j_1$. Indeed, after a convolution of $I(\boldsymbol x)$ by $\psi_{j_1,\theta_1}$, all the information about scales smaller than $2^{j_1}$ is lost, and it is sufficient to consider the modulation of $| I\star \psi_{j_1,\theta_1} |$ by a larger scale $2^{j_2}$. The whole information about the coupling of two scales $j_1$ and $j_2$ is thus contained in $S_2(j_1,\theta_1,j_2,\theta_2)$ for $j_2>j_1$ and for all $\theta_1$ and $\theta_2$. There are then $N_1 = J\cdot \Theta$ coefficients for the $m=1$ layer and $N_2 = J\cdot (J-1)/2 \cdot \Theta^2$ coefficients for the $m=2$ layer. For $J=6$ and $\Theta=8$, which are the values we consider in this paper\footnote{Working mainly with fields of linear sizes $2^8$ pixels, one can hardly compute meaningful statistics on scales larger than or equal to $2^6$ pixels. Also, to decompose the half-circle in $\Theta=8$ angles is a good trade-off between a smooth sampling and the possibility to clearly distinguish between two directions for scales up to $2^5$ pixels on a square lattice.}, it gives $N_1=48$ and $N_2 = 960$. With $N_0=1$, this gives a total of 1009 coefficients.

\begin{figure*}[t]
\begin{center}
\includegraphics[width = 0.99\textwidth]{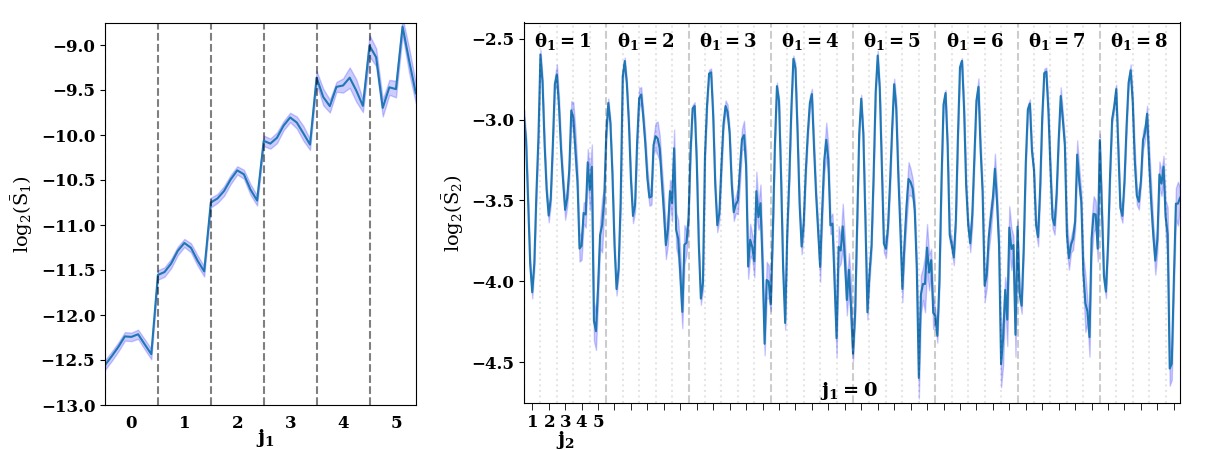}
\vspace{-0.4cm}
\end{center}
 \caption{Logarithms of the normalised scattering coefficients of a 256$\times$256 column density map from an MHD simulation (class 4, see Appendix~\ref{AppendixFlowsStudied}), plotted in a lexicographical order. In the $\log_2 \left[ \bar{S}_1(j_1,\theta_1) \right]$ plot {\it (left)}, $\theta_1$ spans the range 1-8 for each value of $j_1$. In the $\log_2 \left[\bar{S}_2(j_1,\theta_1,j_2,\theta_2) \right]$ plot {\it (right)}, $\theta_2$ spans the range 1-8 for each value of $(j_1=0,\theta_1,j_2)$. The computation of the error bars is explained in Sec.~\ref{PartSecFitVerif}.}
\label{FigScatCoeffEx}
\end{figure*}

Since in simple cases we may expect the scattering coefficients to depend on scales through power laws of $2^{j_1}$ and $2^{j_2}$, it is useful to work with their logarithms, which would lead to linear behaviours as a function of $j_1$ and $j_2$~\citep{sifre2013rotation}. We finally normalise each layer of scattering coefficients by those of the previous layer. We denote $\bar{S}$ these normalised coefficients,
\begin{equation}
\label{EqNormalize1}
\displaystyle{
\log_2 \left[ \bar{S}_1(j_1,\theta_1) \right] = \log_2 \left[ S_1(j_1,\theta_1) \right]- \log_2 \left[ S_0 \right],
}
\end{equation}
and
\begin{multline}
\label{EqNormalize2}
\displaystyle{\log_2 \left[\bar{S}_2(j_1,\theta_1,j_2,\theta_2) \right] = }
\\
\displaystyle{ \log_2 \left[ S_2(j_1,\theta_1,j_2,\theta_2)\right] - \log_2 \left[ S_1(j_1,\theta_1)\right],}
\end{multline}
whereas $\log_2(\bar{S}_0)=\log_2(S_0)$ is unchanged. This normalization separates the dependencies of the different layers~\citep{bruna2015intermittent}, since the $\bar{S}_1$ and $\bar{S}_2$ coefficients are invariant under the multiplication of the field by a constant factor, and the $\bar{S}_2$ coefficients are also invariant under a modification of the spectrum of the field by the action of a linear filter\footnote{This is in fact verified only for linear filters roughly constant on each of the spectral domains sampled by the different wavelets~\citep{bruna2018multiscale}.}. Note that in practice, this normalization is done locally before performing the spatial average (see App.~\ref{AppendixLocalWST} for more details).

As an example, Fig.~\ref{FigScatCoeffEx} shows the logarithms of scattering coefficients [$\log_2(\bar{S}_1)$ on the left and $\log_2(\bar{S}_2)$ on the right] computed from a 256$\times$256 column density map in a simulation of an astrophysical flow (see Appendix~\ref{AppendixPolaris} for more details), with $J=6$ and $\Theta=8$. All the coefficients are plotted as a function of their arguments in lexicographical order: for instance, the first eight $\log_2(\bar{S}_1)$ coefficients are for fixed $j_1=0$ and $\theta_1$ varying from 1 to 8, the next eight $\log_2(\bar{S}_1)$ coefficients are for $j_1=1$ and $\theta_1$ varying from 1 to 8, and so on. In Fig. ~\ref{FigScatCoeffEx} (right), only the $j_1=0$ subset of the $\log_2(\bar{S}_2)$ coefficients is plotted, in a $(j_1,\theta_1,j_2,\theta_2)$ lexicographical order.

\section{The reduced wavelet scattering transform}
\label{PartReducedScatCoeff}

The WST was introduced in data science
with the purpose of characterizing any given field without assuming constraints such as continuity or regularity. In the case of physical fields, some of these constraints may be expected to hold, suggesting possible simplifications. Indeed, the scattering coefficients shown in Fig.~\ref{FigScatCoeffEx} exhibit regular patterns through the angles and scales, for instance the 'stair-like' shape for the $m=1$ coefficients and the oscillatory structure for $m=2$. It should therefore be possible to derive a new statistical description that would somehow factor out these patterns, and thus offer a significant compression of the WST coefficients.

\subsection{Rationale}

We propose such a description, called the Reduced Wavelet Scattering Transform (RWST), obtained by fitting the angular $(\theta_1,\theta_2)$ dependencies of the WST coefficients with a few terms accounting for specific angular modulations. This reduction allows to concentrate the information contained in the $\sim1000$ coefficients of the WST (with $J=6$ and $\Theta=8$) into fewer than 100 reduced coefficients, almost without any loss of information (see Sec.~\ref{PartSecSynthesis} below). Gathering coefficients describing specific angular modulations also allows a simpler and more transparent description, and gives supplementary simplifications in some cases. For example the coefficients describing the statistical anisotropies of a field can be ignored when the latter is in fact statistically isotropic.

\begin{figure*}
\begin{center}
\includegraphics[width = 0.99\textwidth]{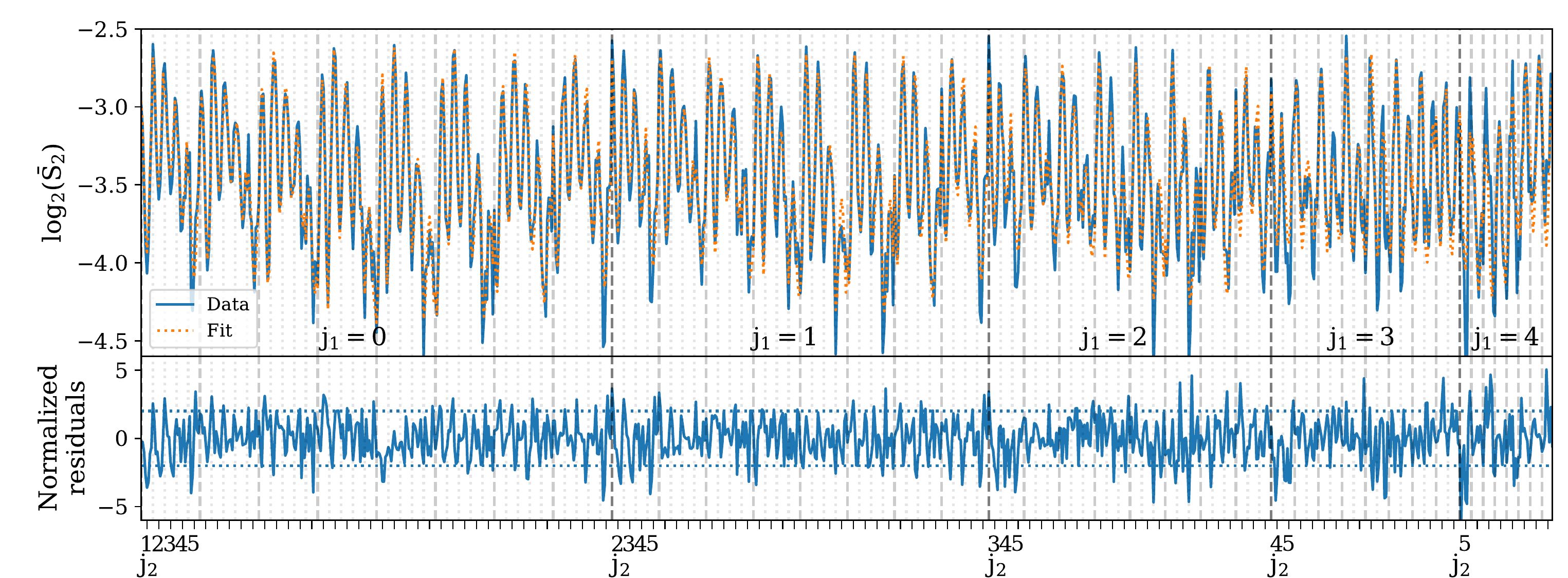}
\vspace{-0.4cm}
\end{center}
 \caption{Logarithms of the normalised $m=2$ scattering coefficients of a 256$\times$256 column density map from an MHD simulation (class 4, see Appendix~\ref{AppendixFlowsStudied}). The coefficients are given in lexicographical order $(j_1,\theta_1,j_2,\theta_2)$. In the top panel, the fit using Eq.~(\ref{EqScatDecomp2}) is shown (dashed orange line) on top of the data (solid blue line). In the bottom panel, we show the residuals normalised to the standard deviation. Dotted lines correspond to $\pm$2$\sigma$.}
\label{FigFITSC2_Turb_1Map}
\end{figure*}

Our modelling of the WST coefficients separates the dependency on angles from that on scales. In this assumption, the logarithms of WST coefficients may formally be written as a sum of terms corresponding to the various possible modulations, 
\begin{equation}
\label{EqGeneralSCRed}
\displaystyle{
\log_2 \left[ \bar{S}_m(\{j_i,\theta_i\}) \right] =\sum_p \hat{S}_m^{p} (\{j_i\}) \cdot  f^{p}_m (\{\theta_i\}),
}
\end{equation}
for $m=1$ and $2$, while $\log_2(\bar{S}_0) = \hat{S}_0$ for $m=0$. In Eq.~\eqref{EqGeneralSCRed}, the $f^{p}_m (\{\theta_i\})$ are given functions of the angles, that may involve some reference angles related to preferred directions for anisotropic fields, and the $\hat{S}_m^{p} (\{j_i\})$ are the reduced scattering coefficients, giving the respective amplitudes of the various angular modulations. 

Before discussing the form of the functions $f^{p}_m (\{\theta_i\})$, it should first be stressed that rotations and scalings, in their infinitesimal versions, can be thought of as small deformations, under which the WST is continuous (see Sec.~\ref{PartWSTProp}). The scattering coefficients should therefore be seen as a discrete sampling, at scales $\{j_i\}$ and angles $\{\theta_i\}$, of a continuous statistical description, rather than as independent descriptors. The same should therefore be true of the RWST description given in Eq.~(\ref{EqGeneralSCRed}).

\subsection{Reduction of the angular dependency}
\label{PartModelMain}

The choice of the functions $f^{p}_m (\{\theta_i\})$ should be guided by general considerations involving periodicities and angle references. For instance, the presence of a statistically preferential direction, such as in images of fluid flows with a mean direction, should manifest itself by a symmetric modulation of the WST response with a $\pi$-periodicity, since the scattering coefficients are themselves $\pi$-periodic\footnote{Indeed, the Morlet wavelets verify $\psi_{j,\vartheta+\pi} = \psi_{j,\vartheta}^*$. Knowing that $I$ is real-valued, the wavelet coefficients $| I \star \psi_{j,\theta}|$ are then $\pi$-periodic.}, and an angle reference either along or perpendicular to the preferential direction. Similarly, a modulation related to some signature of pixelation should be aligned with the lattice and have a $\pi/2$ periodicity (see Appendix~\ref{AppendixDetailsDuFit}). All periodic functions being susceptible to a Fourier series decomposition, we may assume - to first order - the $f^{p}_m$ to be cosine functions. This assumption, as we will see, is generally validated by the successful fits of the different physical fields it allows, as shown in Sec.~\ref{PartSecFitVerif}. 

We first assume that the fields may be statistically anisotropic, but with only one preferential direction at most. A similar approach could be developed in the case of physical phenomena exhibiting several preferential directions. The angular modulations to take into account are expected to be functions of $\theta_1$, $\theta_2$, or of the difference $\theta_1 - \theta_2$, the latter being isotropic since it does not change under a global rotation.

For the $m=1$ layer, the only angular dependency is on $\theta_1$. Following Eq.~\eqref{EqGeneralSCRed} and the discussion above, we write $\log_2 \left(\bar{S}_1\right)$ as the sum of an isotropic {term independent of $\theta_1$, and an anisotropic term proportional to a $\pi$-periodic cosine function of $\theta_1$:
\begin{multline}
\label{EqScatDecomp1}
\displaystyle{
\log_2 \left[ \bar{S}_1 \left(j_1,\theta_1\right)\right] ~ = ~  \hat{S}_1^\text{iso} \left(j_1\right) 
}
\\
\displaystyle{+~  \hat{S}_1^\text{aniso}\left(j_1\right) \cdot \cos\left(\frac{2\pi}{\Theta} ~ \Big[\theta_1-\theta^\text{ref,1}(j_1)\Big]\right),
 }
\end{multline}
where $\theta^{\text{ref,1}}(j_1)$ is a reference angle related to the direction of anisotropy. This angle is a function of the scale $j_1$, and is expected to smoothly vary across the scales\footnote{Although the potential dependency of $\theta^{\text{ref,1}}$ on $j_1$ means that scales and angle dependencies are not completely separate as Eq.~(\ref{EqGeneralSCRed}) suggests, we use this slightly more general form to be able to detect variations of the anisotropy directions across the scales.}. Such a trigonometric function distinguishes a direction from the perpendicular one, but not a direction from its opposite\footnote{Note that in terms of geometrical angles $\vartheta$ associated with the integer labels $\theta$ [see Eq.~\eqref{EqDefTheta}], the cosine function reads $\cos[2(\vartheta_1 - \vartheta^{\text{ref,1}})]$ and is therefore $\pi$-periodic. Note also that the reference angles are fitted as real values in $[0,\pi)$, and not as integers, in order to describe all possible directions. They may also be defined modulo $\pi/2$ by reversing the sign of $\hat{S}_1^\text{aniso}(j_1)$, but this degeneracy can be lifted by enforcing $\hat{S}_1^\text{aniso}\geqslant0$.}. If the field is statistically isotropic, then we expect $\hat{S}_1^{\text{aniso}} \simeq 0$.

\begin{figure*}
\begin{center}
\includegraphics[width = 0.99\textwidth]{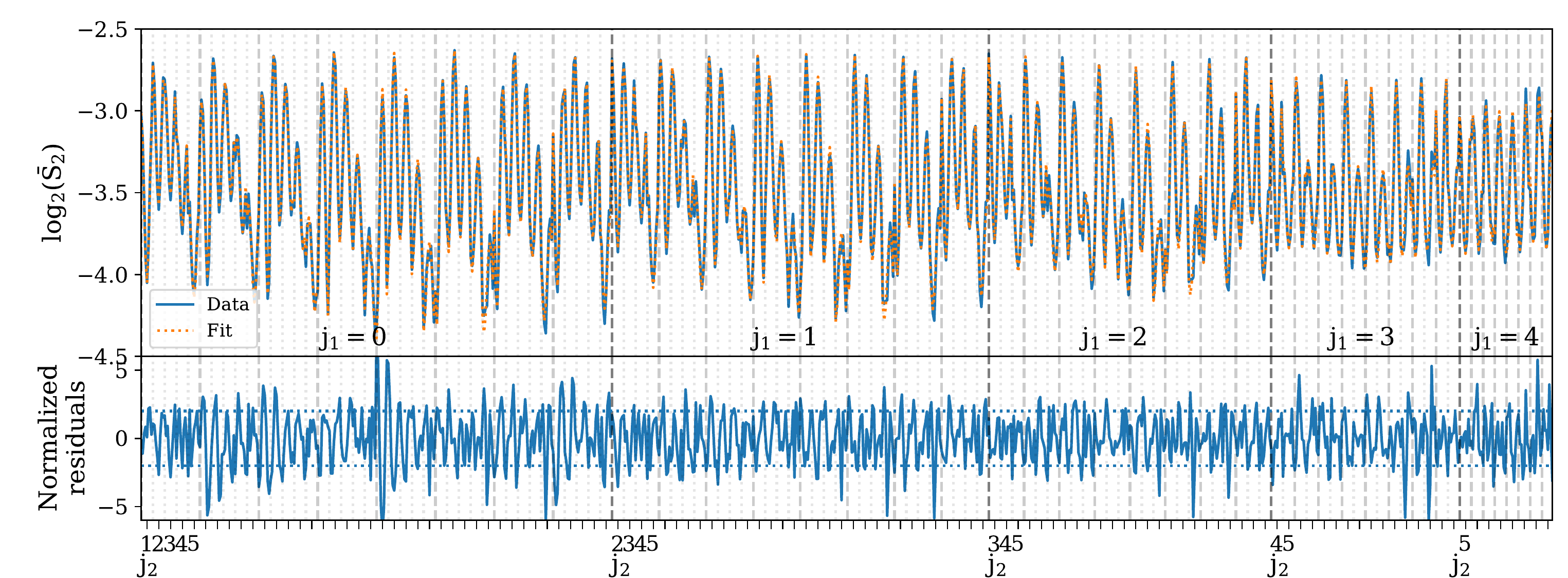}
\vspace{-0.4cm}
\end{center}
 \caption{Logarithms of the normalised $m=2$ scattering coefficients averaged over twenty 256$\times$256 column density maps from an MHD simulation (class 4, see Appendix~\ref{AppendixFlowsStudied}). The coefficients are given in lexicographical order $(j_1,\theta_1,j_2,\theta_2)$. In the top panel, the fit using Eq.~(\ref{EqScatDecomp2}) plus additional terms  discussed in Appendix~\ref{AppendixDetailsDuFit} is shown (dashed orange line) on top of the data (solid blue line). The additional terms  that have been taken into account in this fit are a lattice signature and a $\pi/2$ harmonic for the $\hat{S}_2^{\text{iso,2}}$ term. In the bottom panel, we show the residuals normalised to the standard deviation. Dotted lines correspond to $\pm2\sigma$.}
\label{FigFITSC2_Turb_20Map}
\end{figure*}

For the $m=2$ layer, we consider the following four-term decomposition, with two isotropic and two anisotropic terms,
\begin{align}
\label{EqScatDecomp2}
\log_2 &  \big[\bar{S}_2 ( j_1,\theta_1,j_2,\theta_2)  \big]~ =  ~ \hat{S}_2^\text{iso,1}\left(j_1,j_2\right) 
\nonumber
\\
~ & \displaystyle{
+~  \hat{S}_2^\text{iso,2}\left(j_1,j_2\right) \cdot \cos\left(\frac{2\pi}{\Theta} ~ \Big[\theta_1-\theta_2\Big]\right)
 } 
 \nonumber
 \\
~ &  \displaystyle{
+~   \hat{S}_2^\text{aniso,1}\left(j_1,j_2\right) \cdot\cos\left(\frac{2\pi}{\Theta} ~ \Big[\theta_1-\theta^\text{ref,2}(j_1,j_2)\Big]\right)
}
\nonumber
\\
~ & \displaystyle{
+~ \hat{S}_2^\text{aniso,2}\left(j_1,j_2\right) \cdot \cos\left(\frac{2\pi}{\Theta} ~ \Big[\theta_2-\theta^\text{ref,2}(j_1,j_2)\Big]\right)
 } .
\end{align}
All the cosine functions in this equation are $\pi$-periodic, as in the $m=1$ case. We ignore a potential reference angle difference for the $\hat{S}_2^\text{iso,2}$ term and assume the same $\theta^\text{ref,2}(j_1,j_2)$ reference angle for the two anisotropic terms, further imposing that it should be close to $\theta^\text{ref,1}(j_1)$.

Our statistical description thus consists in eight functions ($\hat{S}_1^\text{iso}$, $\hat{S}_1^\text{aniso}$, $\theta^\text{ref,1}$, $\hat{S}_2^\text{iso,1}$, $\hat{S}_2^\text{iso,2}$, $\hat{S}_2^\text{aniso,1}$, $\hat{S}_2^\text{aniso,1}$, and $\theta^\text{ref,2}$) that are discretely sampled at scales $j_1$ and $j_2$. These functions form the reduced wavelet scattering transform, and each of them is described by $J$ coefficients for $m=1$, and $J(J-1)/2$ coefficients for $m=2$, in addition to the $m=0$ coefficient, for a total of $(5J+1)J/2+1$ coefficients. This gives for instance 94 coefficients for $J=6$. More precisely, the $48$ WST coefficients of the $m=1$ layer are fitted with $18$ degrees of freedom, while the $960$ WST coefficients of the $m=2$ layer are fitted with $75$ degrees of freedom\footnote{Note that the different components of fixed $j_1$ for $m=1$ and of fixed $(j_1,j_2)$ for $m=2$ can be fitted independently.}.

We investigated the limits of this reduction of the WST coefficients. Our study shows that the modulations given in Eqs.~\eqref{EqScatDecomp1} and~\eqref{EqScatDecomp2} are always largely sufficient to describe the angular dependencies of the scattering coefficients. In other words, this reduction boils down to the scattering coefficients at fixed $j_1$ and $j_2$ being described by the zeroth and first harmonics in $\theta_1$ and $\theta_2$, with a very good approximation. Indeed, higher harmonics of those angular modulations are not detectable when working with a single map, and are detected at a very small level when working with a set of 20 independent maps for a given process (see below). It was also possible to identify minor modulations associated to potential signatures of the lattice at the smallest scales. These terms, that allow to better evaluate the limits of the RWST, are discussed in Appendix~\ref{AppendixDetailsDuFit}. Note however that in any case, their addition does not essentially modify the values of the reduced scattering coefficients obtained by fitting Eqs.~\eqref{EqScatDecomp1} and~\eqref{EqScatDecomp2}.

\begin{figure*}[t]
\begin{center}
\includegraphics[width = 1\textwidth]{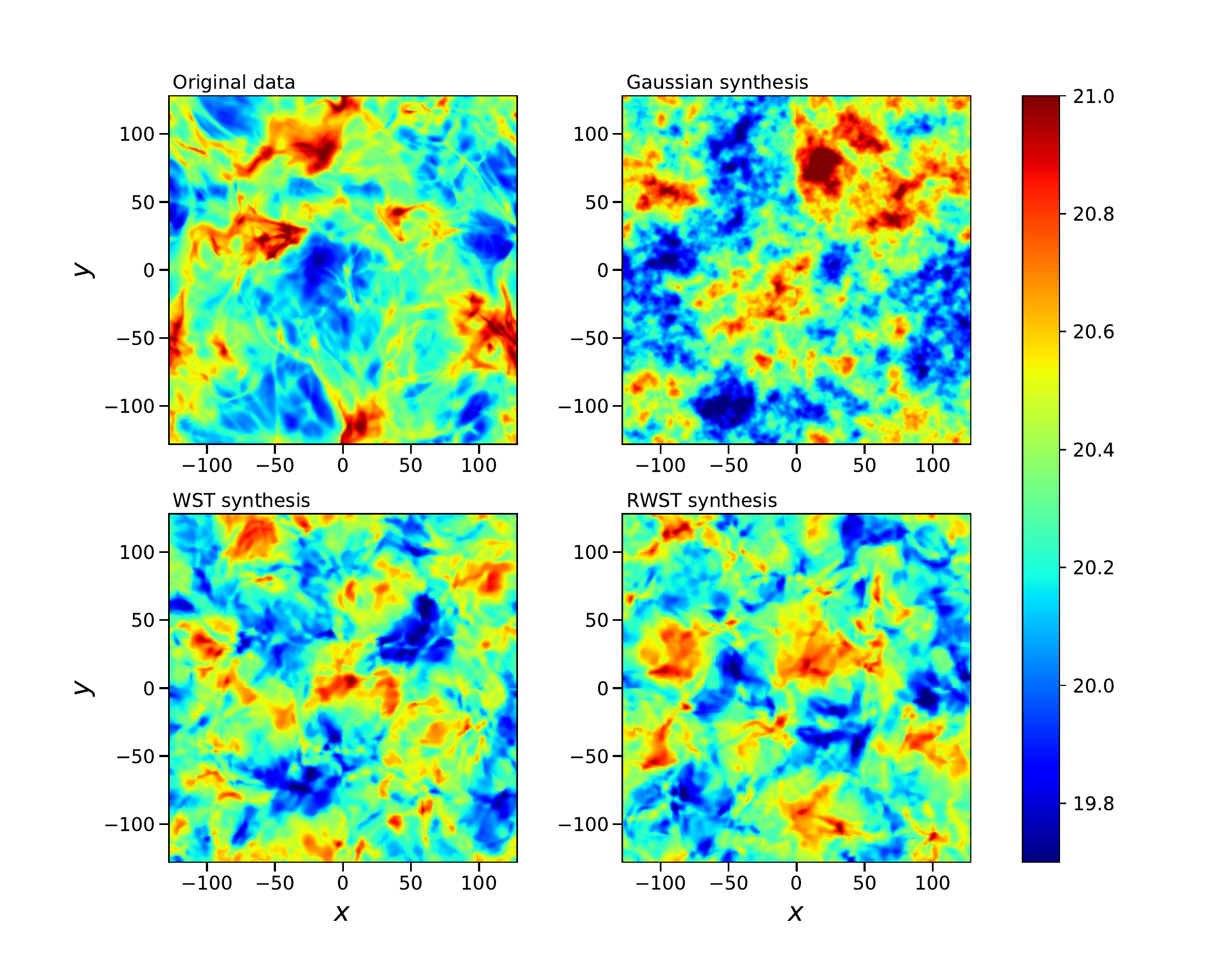}
\vspace{-0.7cm}
\end{center}
 \caption{{\it Top left:} Example of a column density map, in logarithmic scale, from a simulation of interstellar MHD turbulence (class 5, see Appendix~\ref{AppendixFlowsStudied}). {\it Top right:} Synthetic Gaussian random field with the same power spectrum. {\it Bottom left:} Synthetic random field with the same $m=0$, $m=1$, and $m=2$ WST coefficients. {\it Bottom right:} Synthetic random field with the same $m=0$, $m=1$, and $m=2$ RWST coefficients.}
\label{FigSyntheses}
\end{figure*}

\subsection{Test cases}
\label{PartSecTestCases}

The  fit of the angular dependency of the WST coefficients and the associated reduction have been tested on various fields. These are presented in detail in Appendix~\ref{AppendixFlowsStudied}, but we give here a short description of each of them for easy reference.

The first type of field used are realizations of fractional Brownian motions~\citep[fBm,][]{stutzki1998fractal}, Gaussian random fields with power-law power spectra characterised by a Hurst exponent $H\in[0,1]$. We explore the range $H=0.1$ to $H=0.9$ in steps of $0.1$. For each value of $H$, we use 20 different random realizations over a $256\times256$ grid. In the following, we may refer to fits using a single map or using the ensemble of 20 maps.

The second type of field used are $256\times256$ gas column density maps $N_\mathrm{H}$ obtained from numerical simulations of magnetised turbulent astrophysical flows~\citep{Iffrig17}. There are nine classes of such maps, labelled from 1 to 9, with varying intensities of the magnetic field and of the turbulent velocity forcing (see Table~\ref{Tab:MHDclasses}). For each class, we use 20 independent maps. Similarly to the fBm case, in the following, we may refer to fits using a single map or using the ensemble of 20 maps.

The third and last field used is an observation of the dust continuum thermal emission in the Polaris Flare molecular cloud~\citep{miville2010herschel} with the {\it Herschel} satellite~\citep{pilbratt2010herschel,griffin2010herschel}. Unlike the first two types of field, the statistics of this field are likely not homogeneous. We roughly addressed this limitation by using a local WST (see Appendix~\ref{AppendixLocalWST}) and clustering the data into four sub-regions ({\it clusters}) based on the local WST coefficients, although this process and the chosen number of clusters are somewhat arbitrary. 

\subsection{Goodness of the fits}
\label{PartSecFitVerif}

The local computation of the WST coefficients (Appendix~\ref{AppendixLocalWST}) is also instrumental in evaluating the goodness of the fits. The statistical dispersion of these local WST coefficients over each map and over the different realizations allows to estimate the empirical variance of the global WST coefficients, and in turn their uncertainties. As we work with non-Gaussian processes for which no analytic variance estimation is available, this method is currently the only one we have at hand to estimate these uncertainties. One should however keep in mind that this method has a notable flaw. Indeed, while the empirical estimate of the variance of the WST coefficients has to converge to its expected value when using a large enough number of samples, such an empirical estimate can be of poor quality when this convergence is not achieved. 

From our empirical study, we assess that the sampling performed in a single $256\times256$ map only gives well determined uncertainties only for scales $j\leqslant3$, while it is necessary to sample on 20 such maps to correctly determine the uncertainties for scales up to $j=5$. This result can be seen for instance  in Fig.~\ref{FigScatCoeffEx}, where scattering coefficients obtained in a single map are given, as well as their estimated uncertainties. One can for example see in both panels that the uncertainties on the coefficients are underestimated for $j\geqslant4$. This is particularly visible for $j=5$.

The fits of the angular dependencies given in Eqs.~\eqref{EqScatDecomp1} and~\eqref{EqScatDecomp2} were performed taking these statistical uncertainties into account, yielding a standard $\chi^2_\text{red}$ using a diagonal covariance matrix. We chose not to include complete covariance matrices because we cannot properly estimate them on a single map. This implies, in addition to the previous discussion on the statistical uncertainties of the scattering coefficients, that these $\chi^2_\text{red}$ are only indicative of the goodness of the different fits. This is something to be improved upon in future works.

Performing these fits separately for each of the fBm and MHD simulation maps at our disposal yields as many $\chi^2_\text{red}$ values as there are realizations, i.e, $2\times9\times20=360$. Over all of these, we have found similar results, that boil down to an average $\chi^2_\text{red}$ of 3.5 with a dispersion of 0.6 for the $m=2$ coefficients. Such a fit is shown in Fig.~\ref{FigFITSC2_Turb_1Map}. Similar results are also obtained when fitting the combined data from all twenty realizations for each class of MHD simulation or $H$ exponent of the fBm field, provided that the minor additional terms described in App.~\ref{AppendixDetailsDuFit} are included. Such a fit is shown in Fig.~\ref{FigFITSC2_Turb_20Map}.

The goodnesses of the fits are somewhat lower in the case of the {\it Herschel} observations of the Polaris Flare. For the four sub-regions of the cloud discussed in Appendix~\ref{AppendixPolaris}, we obtain $\chi^2_\text{red}$ values with a mean of 7.2 and a standard deviation of 3.3 for the $m=2$ fits. We however observe that the smallest scales are very noisy in these fields. Indeed, performing the same fits while excluding the $(j_1=0,j_2=1)$ scale, we obtain $\chi^2_\text{red}$ values with a mean of 4.8 and a standard deviation of 1.2. We consider these values to be satisfactory, given the heterogeneity of the statistics across the field of view and the crudeness of the clustering approach we have used to address it.

These results validate our reduction of the WST to the RWST, and show the wide range of applicability of this new statistical description. 
Note however that we expect this reduction to be efficient on physical fields only. We tested this hypothesis by fitting the angular dependency of the WST coefficients obtained on the image of a brick wall from the UIUC data base~\citep{agarwal2004learning}.We obtained large $\chi^2_\text{red}$ values, with a mean around 450 for $m=2$, because the angular dependencies are in this case not amenable to smooth trigonometric functions.


\subsection{Syntheses}
\label{PartSecSynthesis}

The efficiency of the RWST as a means to capture the essential statistics of a complex field can be assessed through our ability, starting from the RWST coefficients, to build synthetic fields that are visually similar to the original data. Although this is not an absolute criterion, such a visual comparison is a widespread indicator in data science, among others such as the ability to regress physical parameters, or to achieve high rates of success in classification tests. We also note that the power spectrum does not generally pass this test for non-Gaussian fields, which is exemplified by the fact that starting from a highly non-Gaussian image and synthesizing a field that has the same power spectrum but with random Fourier phases completely destroys the structure in the image.

The synthetic fields are constructed from a set of target WST coefficients. These may be obtained directly from the field to mimic, or from its RWST coefficients, using Eqs.~\eqref{EqScatDecomp1} and~\eqref{EqScatDecomp2}. Starting from a white Gaussian noise map in order to ensure high randomness, the pixel values are iteratively modified through a gradient descent method to obtain the expected WST coefficients [see~\cite{bruna2018multiscale} for more details]. We show such syntheses in Fig.~\ref{FigSyntheses}, starting from an original data set that consists of 20 column density maps from a simulation of interstellar MHD turbulence (class 5, see Appendix~\ref{AppendixFlowsStudied}). One of these maps is shown in the top left panel. Three different syntheses are performed based on these original data: a synthetic Gaussian field with the same power spectrum ({\it top right}), a synthetic field with the same WST coefficients ({\it bottom left}), and a synthetic field with the same RWST coefficients ({\it bottom right}). These syntheses are done with the maximum scale $J=6$ we use in this paper. This implies that the structures larger than $2^5=32$ pixels are not properly synthesised, including large scales modulations as well as the long and thin filamentary structures. Further work is needed to include the largest scales in the synthesis algorithm.

\begin{figure*}[t]
\begin{center}
\includegraphics[width = 0.95\textwidth]{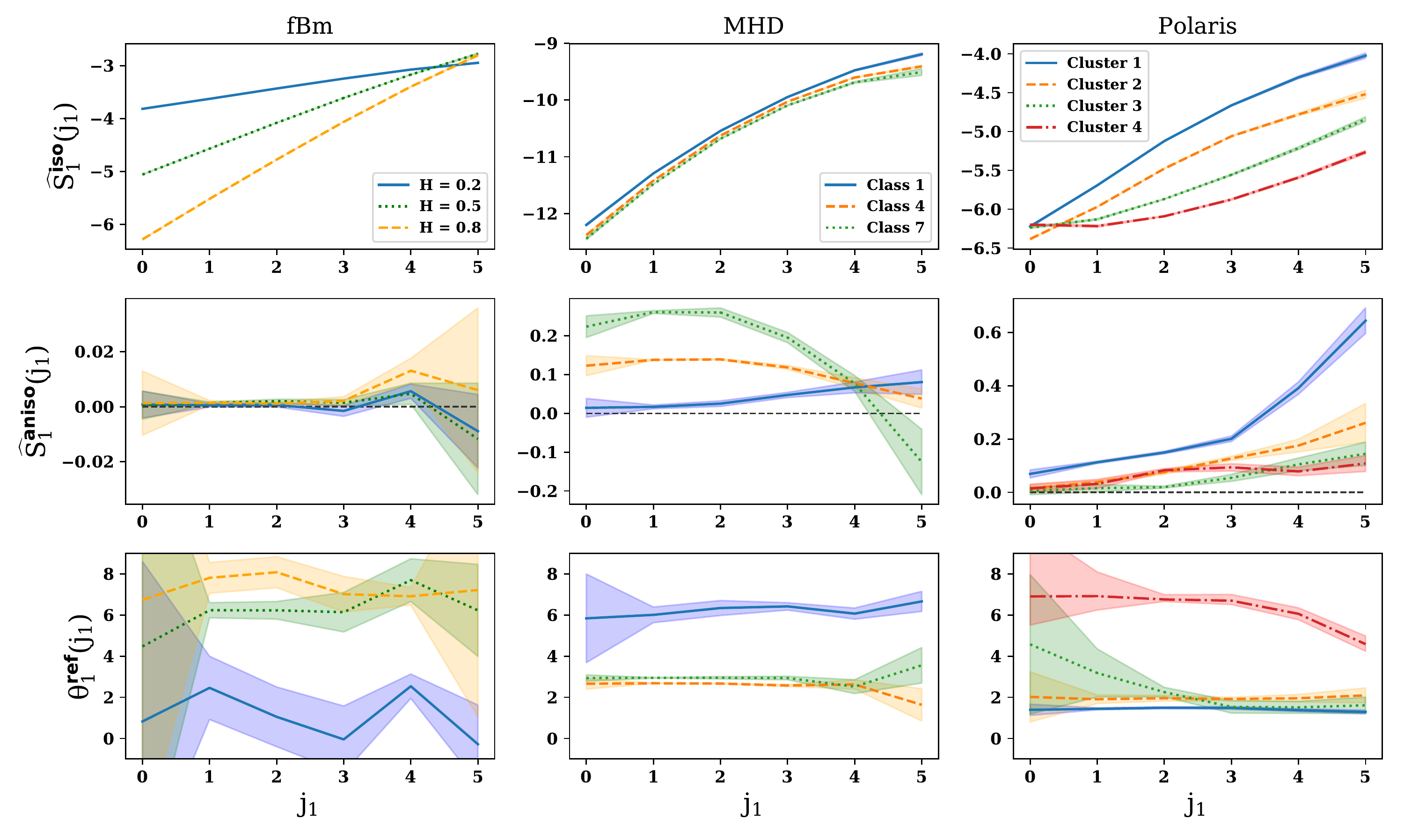}
\vspace{-0.7cm}
\end{center}
 \caption{Plots of $\hat{S}_1^\text{iso}(j_1)$ (\textit{top row}), $\hat{S}_1^\text{aniso}(j_1)$ (\textit{middle row}), and $\theta^\text{ref,1}(j_1)$ (\textit{bottom row}). The first column shows the case of three fBm processes with different Hurst exponents, the second column three MHD simulations with different physical parameters, and the third column the four clusters of the {\it Herschel} Polaris Flare observation.}
\label{FigRSC_S1}
\end{figure*}

Nevertheless, both of the syntheses based on scattering coefficients provide a better agreement with the original image than does the Gaussian field. More importantly, the RWST-based synthesis is at least as good as the WST-based one, showing that the dimensionality reduction (from $\sim$1000 to fewer than 100 coefficients, and even fewer than 50 for isotropic processes) leads to no significant loss of statistical information. Other similar RWST syntheses are given in Appendix~\ref{AppendixCompleteSets} (Fig.~\ref{FigSynthFull}). The syntheses of fBm processes and MHD simulations shown there also display good agreement with the original data, except at the largest scales, as already discussed. This shows the efficiency of our angular reduction, and the ability of the RWST to characterise in a very compact form a significant part of the relevant statistical information about physical fields with homogeneous statistics.

In the same Fig.~\ref{FigSynthFull}, we also present syntheses of fields using the RWST coefficients computed on clusters of the Polaris Flare field (see Appendix~\ref{AppendixPolaris}). In this case, the syntheses are hampered by the heterogeneous statistics of the original data, as can be seen mainly for cluster 2. Improving this is a direction for future work. Nevertheless, we find present syntheses of sub-regions of the Polaris Flare to be in reasonably good agreement with the original data, and believe that they show the applicability of the RWST to observational data as well as simulations.


\section{Interpretation of the RWST components}
\label{PartVariousTermsDiscussion} 
This section examines the different components of the RWST, and proposes physical interpretations for them, using as examples the three different types of fields introduced in Sec.~\ref{PartSecTestCases} (see also Appendix~\ref{AppendixFlowsStudied}). 

\subsection{Introduction}
\label{PartRWSTDiscussionIntro}

For the fBm processes as well as the column density maps from MHD simulations, the RWST coefficients plotted in this section have been obtained from WST coefficients averaged over the 20 independent maps for each class. For the {\it Herschel} observation, the RWST coefficients are calculated from the WST coefficients averaged over each of the four clusters. To support our physical interpretations, we have explored the full range of parameters for each type of field, varying the Hurst exponent of fBm processes, the physical parameters of the MHD simulations, and the cluster of the Polaris Flare {\it Herschel} observation. In addition to the plots shown in this section, we also refer to several additional plots of RWST coefficients given in Appendix~\ref{AppendixCompleteSets} that are helpful to discuss these explorations of the parameter spaces. 

In the plots discussed, the RWST coefficients for $m=1$ ($\hat{S}_1^\text{iso}$, $\hat{S}_1^\text{aniso}$, and $\theta^\text{ref,1}$) are of course plotted as functions of $j_1$, and the $m=2$ coefficients ($\hat{S}_2^\text{iso,1}$, $\hat{S}_2^\text{iso,2}$, $\hat{S}_2^\text{aniso,1}$, $\hat{S}_2^\text{aniso,2}$, and $\theta^\text{ref,2}$) are plotted as different functions of $j_2$ for fixed $j_1$. Since $j_2>j_1$, the number of points varies from one curve to another. All the plots include associated $1\sigma$ uncertainties that are obtained by propagating the statistical uncertainties of the WST coefficients through the fitting process\footnote{This means that they have the same flaws as the uncertainties of the initial scattering coefficients, and are probably underestimated for the $j\geqslant4$ scales.}. The smoothness of the variations of the RWST coefficients across the scales corroborates our understanding that these are discrete samplings of underlying smooth functions. 

\subsection{Overview of the different terms}

\paragraph{Isotropic $\hat{S}_1^\text{iso}$ component}
~  \vspace{0.15cm}\newline
The $m=1$ isotropic term $\hat{S}_1^\text{iso}$ describes how the amplitude of the field is distributed across the different scales $2^{j_1}$. Various examples\footnote{Note that for the fBm, we did not normalise the $m=1$ coefficients by $S_0$, as described in Eq.~(\ref{EqNormalize1}), because the fBm processes we consider have zero mean.} of this term are given in Fig.~\ref{FigRSC_S1} ({\it top row}). Other examples are shown in Fig.~\ref{FigS1_Full}, obtained by including the additional terms detailed in Appendix~\ref{AppendixDetailsDuFit}, which, one can see by comparing these two figures, do not change the $\hat{S}_1^\text{iso}$ coefficients appreciably. For scale-invariant processes, these coefficients are expected to be linear functions of the scale, $\hat{S}_1^\text{iso}(j_1) \propto \hat{S}_1^0 + j_1 H$ with $H$ the Hurst exponent~\citep{bruna2015intermittent}. 

\paragraph{Anisotropic $\hat{S}_1^\text{aniso}$ component and reference angle $\theta^\text{ref,1}$}
~  \vspace{0.15cm}\newline
The $m=1$ anisotropic term $\hat{S}_1^\text{aniso}$ describes the angular modulation of the WST coefficients for anisotropic fields, with an extremum at the preferential direction $\theta^\text{ref,1}$,
\begin{equation}
\label{EqRWSTAngCont}
\cos\Big(\frac{2\pi}{\Theta}\big[\theta_1-\theta^\text{ref,1}(j_1)\big]\Big) = \cos\Big(2\big[\vartheta_1 - \vartheta^{\text{ref,1}}(j_1)\big]\Big),
\end{equation}
As already mentioned, $\theta^\text{ref,1}$ can be uniquely defined by imposing $\hat{S}_1^\text{aniso}\geqslant0$. Examples of $\hat{S}_1^\text{aniso}$ coefficients are given in Figs.~\ref{FigRSC_S1} ({\it middle row}) and~\ref{FigS1_Full}. For isotropic fields, we expect $\hat{S}_1^\text{aniso}\simeq 0$ and the uncertainty on $\theta^\text{ref,1}$ should be large, for the same reason that the phase of a very low amplitude complex number is poorly determined. For anisotropic fields, $\hat{S}_1^\text{aniso}$ should be non-zero and $\theta^\text{ref,1}$ should be well defined. It is noticeable that in this case, the $\theta^\text{ref,1}$ angle often has almost constant values over all the scales, which strengthens the interpretation that we are indeed probing a particular direction of anisotropy.

\paragraph{Isotropic $\hat{S}_2^\text{iso,1}$ component}
~  \vspace{0.15cm}\newline
The first $m=2$ isotropic term $\hat{S}_2^\text{iso,1}$ describes at which level the $2^{j_1}$ scales are modulated at the larger $2^{j_2}$ scale. In other words, it describes the couplings between scales. Some examples of this term are given in Fig.~\ref{FigRSC_S2_Iso} ({\it top row}), and others in Figs.~\ref{FigS2_Full_fBm},~\ref{FigS2_Full_Turb}, and~\ref{FigS2_Full_Pol} ({\it first column}). We expect this term to depend on $j_2-j_1$ only for scale-invariant fields, since the modulation of a first scale by a second scale then solely depends on the ratio between the two. We also expect this term to decrease as $j_2-j_1$ increases, and this decrease to be all the steeper for fields where scales are only loosely coupled, and shallower in the case of fields with a strong nonlinear behaviour. Note that this this property is related to the notion of intermittency\footnote{In~\cite{bruna2015intermittent}, this is defined as the occurrence of randomly distributed bursts of transient structures at multiple scales. This may be different from the physical notion of intermittency in studies of turbulent flows.} in random fields~\citep{bruna2015intermittent}.

\paragraph{Isotropic $\hat{S}_2^\text{iso,2}$ component}
~  \vspace{0.15cm}\newline
The second $m=2$ isotropic term $\hat{S}_2^\text{iso,2}$ describes an angular modulation of the $m=2$ WST coefficients of the form
\begin{equation}
\cos\Big(\frac{2\pi}{\Theta}\big[\theta_1-\theta_2\big]\Big) = \cos\Big(2\big[\vartheta_1 - \vartheta_2\big]\Big).
\end{equation}
This term quantifies whether, after the filtering of the field at the $(j_1,\theta_1)$ oriented scale, it is more probable to have, at a given $j_2$, a modulation in the same direction ($\theta_2=\theta_1$), in which case $\hat{S}_2^\text{iso,2}>0$, or in the perpendicular direction, in which case $\hat{S}_2^\text{iso,2}<0$. Some examples of $\hat{S}_2^\text{iso,2}$ coefficients are given in Fig.~\ref{FigRSC_S2_Iso} ({\it bottom row}), and others in Figs.~\ref{FigS2_Full_fBm},~\ref{FigS2_Full_Turb}, and~\ref{FigS2_Full_Pol} ({\it second column}). 

Our understanding of these coefficients is that they signal the presence of structures such as filaments in the field. Indeed, in this case, we expect small scale oscillations to be aligned over larger scales along the different filaments, leading to $\hat{S}_2^\text{iso,2}$ coefficients that do not vanish even at large $j_2-j_1$. One can for example see that all these coefficients quickly converge to zero for large $j_2-j_1$ for the fBm processes, which have very few structure, while they rather converge to a constant value for the filamentary MHD simulations. It is also interesting to see that they indicate an increasing presence of structure from the most diffuse (cluster 4) to the denser (cluster 1) areas of the Polaris cloud, which is in agreement with our expectations.

\paragraph{Anisotropic $\hat{S}_2^\text{aniso,1}$ and $\hat{S}_2^\text{aniso,2}$ components, and reference angle $\theta^\text{ref,2}$}
~  \vspace{0.15cm}\newline
The two $m=2$ anisotropic terms $\hat{S}_2^\text{aniso,1}$ and $\hat{S}_2^\text{aniso,2}$ both describe an angular modulation similar to the one given in Eq.~\eqref{EqRWSTAngCont}, and therefore characterise the anisotropy of the field, but with a finer scale dependency. Examples of these coefficients and reference angle are given in Figs.~\ref{FigRSC_S2_Aniso},~\ref{FigS2_Full_fBm},~\ref{FigS2_Full_Turb}, and~\ref{FigS2_Full_Pol}. Similarly to $\hat{S}_1^\text{aniso}$, we expect $\hat{S}_2^\text{aniso,1}$ and $\hat{S}_2^\text{aniso,2}$ to vanish for statistically isotropic fields, in which case the uncertainty on $\theta^\text{ref,2}$ should be large. For the anisotropic fields, it is striking to note that the levels as well as the direction of anisotropy given by the $m=1$ and $m=2$ reduced scattering coefficients are similar, see for instance Figs.~\ref{FigRSC_S1} and~\ref{FigRSC_S2_Aniso}. This confirms our identification of the physical meaning of these terms.

\begin{figure*}[t]
\begin{center}
\includegraphics[width = 0.95\textwidth]{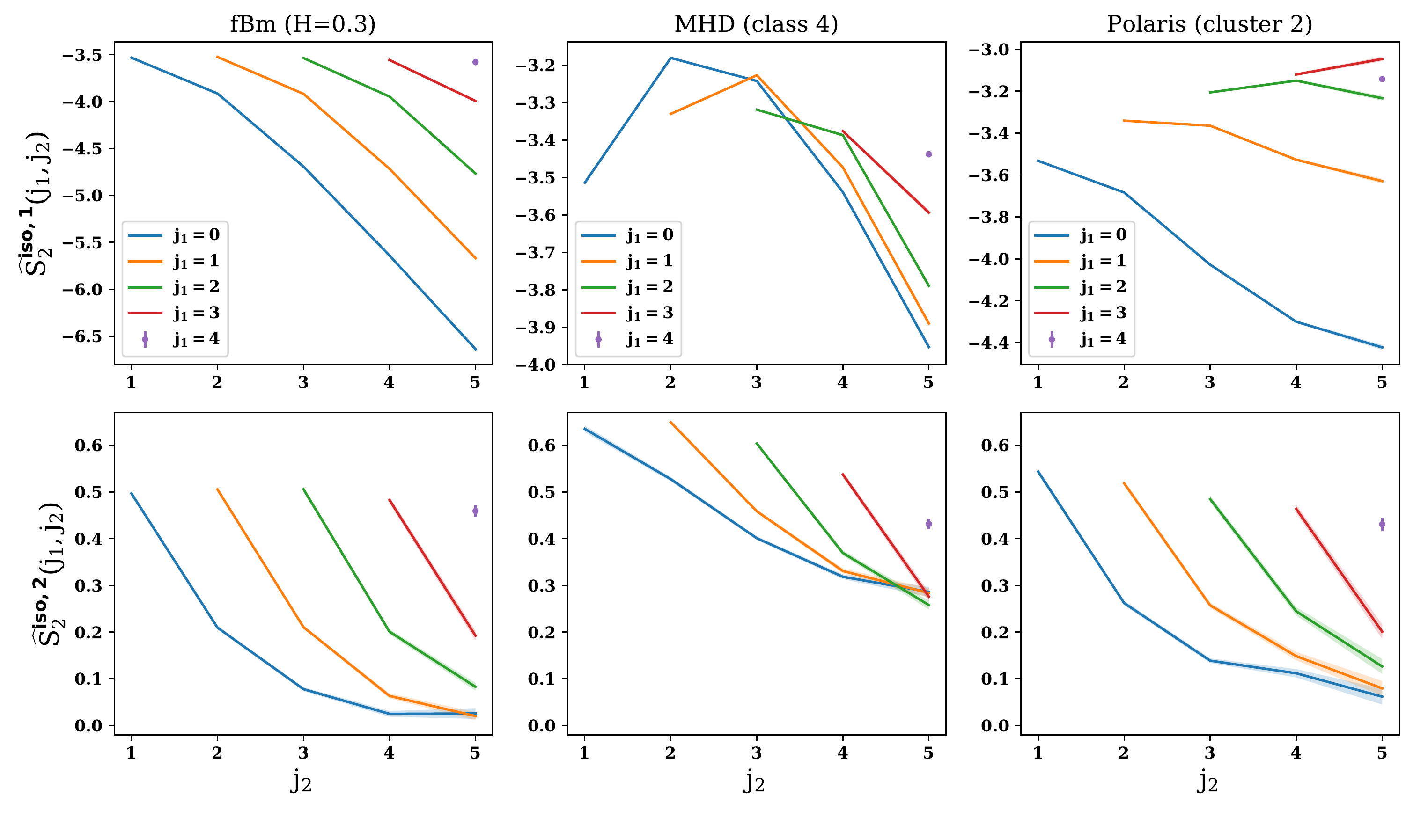}
\vspace{-0.7cm}
\end{center}
 \caption{Plots of $\hat{S}_2^\text{iso,1}(j_1,j_2)$ (\textit{top row}) and $\hat{S}_2^\text{iso,2}(j_1,j_2)$ (\textit{bottom row}). The first column shows the case of a $H=0.3$ fBm process, the second column a MHD simulation (class 4), and the third column the first cluster of the {\it Herschel} Polaris Flare observation. Each curve corresponds to a fixed $j_1$ value, and $j_2$ values ranging from $j_1+1$ to 5.}
\label{FigRSC_S2_Iso}
\end{figure*}

\subsection{Physical interpretations on the various fields}

\paragraph{Scale invariance}
~  \vspace{0.15cm}\newline
The signposts of scale invariance are unsurprisingly most apparent for fractional Brownian motion fields, whose power spectra display power-law scalings. Indeed, for these fields we find that $\hat{S}_1^\text{iso}$ is a linear function of $j_1$ with a slope proportional to $H$ (Fig.~\ref{FigRSC_S1}, {\it top left}), while $\hat{S}_2^\text{iso,1}$ is a function of $j_2-j_1$ only, since the curves for different $j_1$ in Fig.~\ref{FigRSC_S2_Iso} ({\it top left}) come together when plotted as functions of $j_2-j_1$. The same is true of $\hat{S}_2^\text{iso,2}$ (Fig.~\ref{FigRSC_S2_Iso}, {\it bottom left}). 

For the gas column density maps from MHD simulations, we do not observe such a scale-invariant behaviour in the range of scales that is sampled (see the $\hat{S}_1^\text{iso}$ isotropic term in Fig.~\ref{FigRSC_S1}, {\it top centre}). This is not unexpected, since the energy injection in the simulations is itself not scale-invariant. There is a hint of a scale-invariant behaviour at small scales, but these is over too short a range to be meaningful~\citep{frisch1995turbulence}. On the other hand, the $\hat{S}_2^\text{iso,1}$ (Fig.~\ref{FigRSC_S2_Iso}, {\it top centre}) and $\hat{S}_2^\text{iso,2}$ (Fig.~\ref{FigRSC_S2_Iso}, {\it bottom centre}) terms are not a function of $j_2-j_1$ only, indicating that the couplings between scales are not scale-invariant.

In the Polaris Flare {\it Herschel} map, we observe a behaviour that is similar to the MHD simulation maps for the most intense region (cluster 1, Fig.~\ref{FigRSC_S1}, {\it top right}), but also a flattening at small scales in the most diffuse regions\footnote{Recall that the $m=1$ coefficients are normalised by the $m=0$ ones [Eq.~\eqref{EqNormalize1}], which precludes a direct comparison of the $\hat{S}_1^\text{iso}$ values.} (clusters 3 and 4). This may be an effect of noise or of the Cosmic Infrared Background (CIB) fluctuations~\citep{puget1996cib,lagache2005cib,viero2013hermes} beginning to stand out. We also note that the third and four clusters seem to have $\hat{S}_2^\text{iso,1}$ and $\hat{S}_2^\text{iso,2}$ coefficients similar to the scale-invariant fBm ones at the smallest scales, which strengthens this observation.

\begin{figure*}[t]
\begin{center}
\includegraphics[width = 0.95\textwidth]{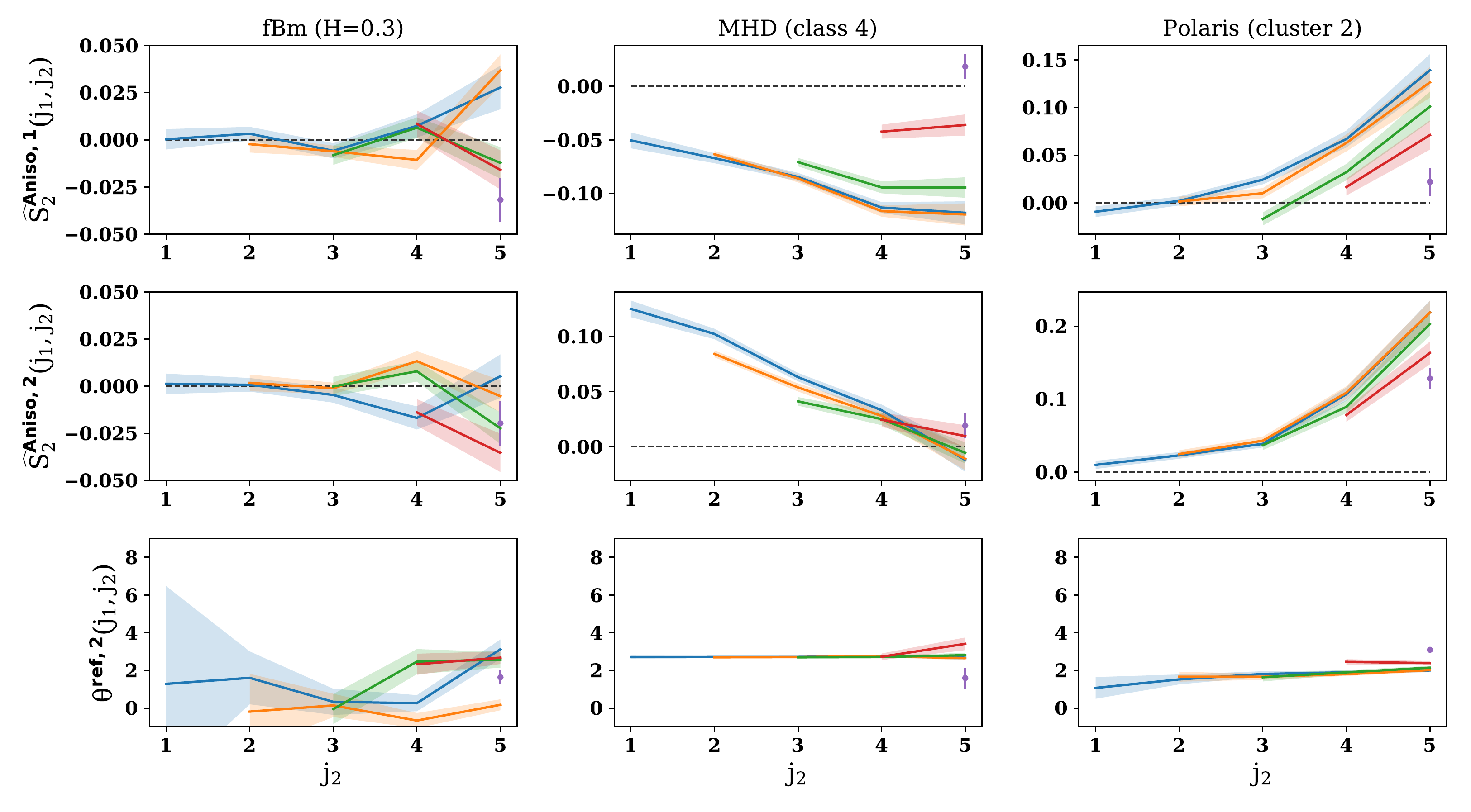}
\vspace{-0.7cm}
\end{center}
 \caption{Plots of $\hat{S}_2^\text{aniso,1}(j_1,j_2)$, $\hat{S}_2^\text{aniso,2}(j_1,j_2)$, and associated $\theta^\text{ref,2}(j_1,j_2)$ terms of the RWST. The first column describes a $H=0.3$ fBm processes, the second a MHD simulations of the fourth described class, and the third column the first clustered area of the Herschel Polaris observation (see Appendix~\ref{AppendixFlowsStudied} for more detail). Each curve has a fixed $j_1$ values, following the convention given in Fig.~\ref{FigRSC_S2_Iso}, and $j_2$ values going from $j_1+1$ to 5.}
\label{FigRSC_S2_Aniso}
\end{figure*}

\paragraph{Couplings between scales}
~  \vspace{0.15cm}\newline
The lack of coupling between scales in fractional Brownian motion fields appears in the fast decrease of $\hat{S}_2^\text{iso,1}$ (Fig.~\ref{FigRSC_S2_Iso}, {\it top left}) and of $\hat{S}_2^\text{iso,2}$ to zero (Fig.~\ref{FigRSC_S2_Iso}, {\it bottom left}) for $j_2-j_1 \geqslant 3$. In addition, the fBm processes share the same structure in $\hat{S}_2^\text{iso,1}$ and $\hat{S}_2^\text{iso,2}$ coefficients. Indeed, one can recover these curves from one another by a simple linear dilation of the $j_i$ scale that depends on their $H$ exponents only~(see Fig.~\ref{FigS2_Full_Turb}). This property can be related to the fact that all Gaussian fields have the same $m=2$ WST coefficients in one dimension~\citep{bruna2015intermittent}, thus allowing their identification independently of their power spectrum.

On the contrary, the $\hat{S}_2^\text{iso,1}$ and $\hat{S}_2^\text{iso,2}$ coefficients for MHD simulations cannot be directly mapped from one another (see Fig.~\ref{FigS2_Full_Turb}). However, these terms share similar forms indicating that the gas dynamics in these MHD simulations are computed for a common set of equations. We expect the differences between those patterns to echo the differences of physical parameters. The dynamic range of $\hat{S}_2^\text{iso,1}$ (Fig.~\ref{FigS2_Full_Turb}, {\it first column}) may be used as a measure of the strength of the coupling between scales. We observe that it decreases as the turbulent forcing increases, in the absence of a mean magnetic field (from class 1 to class 3, see Appendix~\ref{AppendixFlowsStudied}), but that this effect is much less marked when the mean magnetic field is strong (from class 7 to class 9). A similar conclusion may be drawn from the comparison of the dynamic ranges of $\hat{S}_2^\text{iso,2}$ (Fig.~\ref{FigS2_Full_Turb}, {\it second column}) for the same classes. In all cases, this decrease is much less steep than in the fBm case, especially for the $\hat{S}_2^\text{iso,1}$ terms, clearly indicating a stronger coupling between scales. 

For the Polaris Flare map, the $\hat{S}_2^\text{iso,1}$ terms signal a systematic decrease of the coupling between scales, from the most dense region (cluster 1) to the most diffuse (cluster 4, see Fig.~\ref{FigS2_Full_Pol}, {\it first column}). The signature in $\hat{S}_2^\text{iso,2}$ (Fig.~\ref{FigS2_Full_Pol}, {\it second column}) is not so clear-cut, but we do observe that for cluster 1, $\hat{S}_2^\text{iso,2}$ does not go to zero at large $j_2-j_1$, while it does for cluster 4. This indicates a stronger nonlinear dynamics in the denser region of the Polaris Flare, as one could expect. 

\paragraph{Statistical isotropy and anisotropy}
~  \vspace{0.15cm}\newline
The statistical isotropy of fBm fields is evidenced by the fact that $\hat{S}_1^\text{aniso}$ (Fig.~\ref{FigRSC_S1}, {\it middle left}), $\hat{S}_2^\text{aniso,1}$ (Fig.~\ref{FigRSC_S2_Aniso}, {\it top left}), and $\hat{S}_2^\text{aniso,2}$ (Fig.~\ref{FigRSC_S2_Aniso}, {\it middle left}) are all compatible with zero within statistical uncertainties, and thus the uncertainties on $\theta^\text{ref,1}$ (Fig.~\ref{FigRSC_S1}, {\it bottom left}) and $\theta^\text{ref,2}$ (Fig.~\ref{FigRSC_S2_Aniso}, {\it bottom left}) are large. It is also interesting to note that the third and fourth clusters of the Polaris Flare also have isotropic signatures at the scales on which we identified a possible contamination by noise of by the CIB (see Figs.~\ref{FigRSC_S1} and~\ref{FigS2_Full_Pol}).

On the contrary, the MHD simulations mostly display signatures of a statistical anisotropy, that is the result of a competition between the mean magnetic field and the turbulent forcing. Indeed, we note that the larger the magnetic field at a given turbulent forcing (from class 1 to 7 in Fig.~\ref{FigRSC_S1}, {\it centre}), the larger the $\hat{S}_1^\text{aniso}$ terms. Exploring these dependencies further, we note that signatures of anisotropy ($\hat{S}_1^\text{aniso}\neq0$) are clear for MHD simulations with a strong mean magnetic field and low turbulent forcing (class 7, in Fig.~\ref{FigS1_Full}, \textit{centre}), but are smaller for simulations with no mean magnetic field (classes 1 and 3) or with a high turbulent forcing even when the magnetic field strength is large (class 9). Quantitatively, the value of $\hat{S}_1^\text{aniso}$ yields levels of anisotropy of $30\%$ at most for the MHD simulation maps. The same conclusion can be drawn from the study of $\hat{S}_2^\text{aniso,1}$ and $\hat{S}_2^\text{aniso,2}$ (Fig~\ref{FigS2_Full_Turb}).

It is interesting to note that small signatures of anisotropy appear even for the simulations without large-scale magnetic fields (Figs.~\ref{FigS1_Full} and~\ref{FigS2_Full_Turb}, classes 1 and 3). They are \emph{a priori} the result of the inherently anisotropic dynamics of MHD flows. It is also worth noting that those self-induced spontaneous anisotropies have different signatures compare to the ones driven by a mean magnetic field. Indeed, $\hat{S}_1^\text{aniso}$ appears to increase with scale for classes 1 and 3, while it decreases for class 7 (Fig.~\ref{FigS1_Full}, {\it centre}). Similar differences of behaviour also appear for $\hat{S}_2^\text{aniso,1}$ and $\hat{S}_2^\text{aniso,2}$. 

In the Polaris Flare map, we also detect signatures of anisotropy in $\hat{S}_1^\text{aniso}$ (Fig.~\ref{FigS1_Full}, {\it bottom centre}). This coefficient increases with scale as in the case of MHD simulations without a mean magnetic field. We note that it also globally increases from the most diffuse region (cluster 4) to the most intense (cluster 1), reaching a $\sim70\%$ level of anisotropy. It is interesting to note that the $\theta^\text{ref,i}$ reference angles are similar for all clusters, except for the most diffuse one (cluster 4), once again singling it out. The $m=2$ anisotropic RWST coefficients $\hat{S}_2^\text{aniso,1}$ and $\hat{S}_2^\text{aniso,2}$ similarly increase with scale (at least for clusters 1 to 3), also in clear contrast to the MHD simulations with mean magnetic fields (see Figs.~\ref{FigS2_Full_Turb} and~\ref{FigS2_Full_Pol}, {\it third and fourth columns}).


\section{Conclusions and perspectives}
\label{PartConclusion}

We have presented  the RWST, a low-dimensionality statistical description of complex structures arising from nonlinear phenomena, in particular interstellar MHD turbulence. This description is built from the WST, a low-variance statistical description of non-Gaussian processes, developed in data science, that encodes long-range interactions through a hierarchical multiscale approach based on the wavelet transform. The WST characterises the textures of 2D images with coefficients that depend on scales and orientations. The RWST provides a reduction of the WST through a fit of its angular modulations, gathering the information into a few  functions that separate isotropic and anisotropic characteristics of the data.

We have applied the RWST to statistically describe and compare fields arising from three processes: fractional Brownian motions, column density maps from numerical simulations of interstellar MHD turbulence, and an observation of the dust thermal emission from an interstellar cloud (the Polaris Flare). Our analysis, performed on these fields, allows us to draw a number of conclusions on the properties of the RWST.

Firstly, the RWST characterises and differentiates processes with a small number of coefficients grouped into a few functions, since each of the 256$\times$256 maps we have analysed is characterised by 94 RWST coefficients grouped into eight functions of the scales. 
The coefficients are statistical descriptors encoding, with reduced variance, moments of order up to four. 
The coefficients derived from independent realizations of
fractional Brownian motions and MHD simulations  are remarkably consistent for any given set of input parameters. For the Polaris Flare, the coefficients vary significantly across the image, but we obtain a satisfactory description 
of the data by splitting the image in four regions with distinct characteristics. \\

Secondly, the RWST coefficients compose a comprehensive statistical model that we use to generate synthetic random fields (Sect.~\ref{PartSecSynthesis}). The
textures of the synthesised images are noticeably similar to that of the input data on scales sufficiently sampled to allow for a statistical description. 
This match illustrates the ability of the RWST coefficients  to capture the multiscale correlations intrinsic to non-Gaussian fields. \\

Thirdly, the RWST coefficients quantify the properties of scale invariance, as well as the degree and direction of anisotropy across the scales, in a given field (Sect.~\ref{PartVariousTermsDiscussion}). 
They also encode non-Gaussian characteristics quantifying the coupling between scales as signatures of nonlinear gas dynamics. Further work is needed to precisely understand how to use the RWST to 
characterise the filamentary structure of the interstellar medium and the intermittency of interstellar turbulence. \\

Finally, the RWST project data into a space of reduced dimensionality where observations of the interstellar medium may be compared with numerical simulations in a comprehensive way. 
Such comparisons may contribute to constrain the physical properties of interstellar MHD turbulence. The results presented in Sec.~\ref{PartVariousTermsDiscussion} and Appendix~\ref{AppendixCompleteSets}
illustrate this possibility and point out quantitatively that the numerical simulations used in this paper fail to reproduce the statistical properties observed in the Polaris Flare. 
Further work is needed  to check whether a better match is obtained with more realistic  simulations of interstellar MHD turbulence including the formation of structures through the thermal instability. 

In this paper, the WST and the RWST are applied to images.  It would be interesting to extend this analysis to three-dimensional fields from MHD simulations 
of interstellar turbulence,  and data cubes obtained from spectroscopic observations~\citep[e.g.][]{Hily-Blant08,Blagrave17,pety2017} and Faraday tomography \citep[e.g.][]{Zaroubi15,VanEck19},
to build stationary stochastic models of the turbulent magnetised ISM including intermittency~\citep[e.g.][]{falgarone-et-al-2009,momferratos-et-al-2014}. 

One can also develop 
the WST and RWST to analyse all-sky surveys, such as {\it Planck} data, as a whole, using directional wavelets on the sphere~\citep{demanet2001directional,mcewen-et-al-2007}.
This could open up a path towards generating equivalent random fields to be used for the development of advanced component separation methods. A first example of such an application would be the separation in total intensity between the emission from Galactic dust and the Cosmic Infrared Background~\citep{planck2016-XLVIII}. We also expect to be able to adapt the RWST to fields describing polarised emission (Stokes $I$, $Q$, and $U$) and, from there, to simulate polarised Galactic foregrounds~\citep{vansyngel-et-al-2017,planck2016-l11A}.

\begin{acknowledgements} This research is supported by the Agence Nationale de la Recherche (project BxB: ANR-17-CE31-0022). F. Levrier and F. Boulanger acknowledge support from the European Research Council under the European Union's Horizon 2020 Research \& Innovation Framework Programme (ERC grant agreement ERC-2016-ADG-742719). S. Mallat and S. Zhang acknowledge support by the ERC InvariantClass 320959. This work was also supported by the Programme National 'Physique et Chimie du Milieu Interstellaire' (PCMI) of CNRS/INSU with INC/INP co-funded by CEA and CNES. We gratefully acknowledge fruitful discussions with J.-F. Cardoso, K. Benabed, P. Lesaffre, B. Godard, A. Gusdorf, E. Falgarone, S. Prunet, and J. Neveu. We finally thank the anonymous referee, whose comments helped to significantly improve the manuscript.
\end{acknowledgements}

\appendix
\section{Morlet Wavelets and windowed Fourier transforms}
\label{AppendixMorletWavelets}

Wavelets are waveforms that locally quantify the amplitude of a field in a given range of scales (see for instance~\cite{cohen1995wavelets,van2004wavelets,farge2010multiscale,farge2015wavelet} for a more detailed introduction to wavelets and their application in physics and turbulence). They are constructed by dilating and rotating an initial wavelet, generally called the mother wavelet. Each wavelet samples a given region of the Fourier spectrum of the field under study. 

The wavelets used in this paper are complex Morlet wavelets, also called Gabor wavelets. They are complex analytic wavelets that can efficiently separate the amplitude and phase components of a signal, with a good localization in frequency~\citep{leung2001representing}. They are thus well suited to finely describe the spectrum of a field.
The complex Morlet wavelets are defined from a mother wavelet of parameter $\sigma$, that in the one-dimensional case reads:
\begin{equation}
\label{EqDefMorlet1D}
\psi(x) = \alpha \left( e^{i x} - \beta \right) \cdot e^{- x^2/(2 \sigma^2)}.
\end{equation}
In this equation, $\alpha$ and $\beta=\exp(-\sigma^2/2)$ are normalization factors respectively ensuring that the wavelets have a unit $L_2$ norm and a null average~\citep{ashmead2010morlet}. This mother wavelet is the product of a plane wave of unit wavenumber by a Gaussian window of characteristic size $\sigma$ which localises it. The $\beta$ coefficient can often be neglected when $\sigma>1$. The wavelet $\psi_{j}$ is obtained by a dilation of the mother wavelet:
\begin{equation}
\psi_{j}(x) = 2^{-2j} \psi\left(2^{-j} x\right).
\end{equation}
Such a wavelet and its Fourier transform are plotted in Fig.~\ref{FigWavelet1D}. Neglecting the $\beta$ term in a first approximation, the Fourier transform of the mother wavelet is a Gaussian window of width proportional to $1/\sigma$ and centred on the unit wavenumber $k_\psi = 1$. The Fourier transform of the $\psi_{j}$ wavelet is thus centred on the $2^{-j}$ wavenumber and has a bandwith proportional to $(2^j\sigma)^{-1}$. Thus, convolving a given field with such a wavelet corresponds to bandpass filtering in which the passband is defined by the Fourier transform of the wavelet. As this is done locally, this convolution yields the local level of the signal filtered by the wavelet.

\begin{figure}[h!]
\begin{center}
\includegraphics[width = 0.49\textwidth]{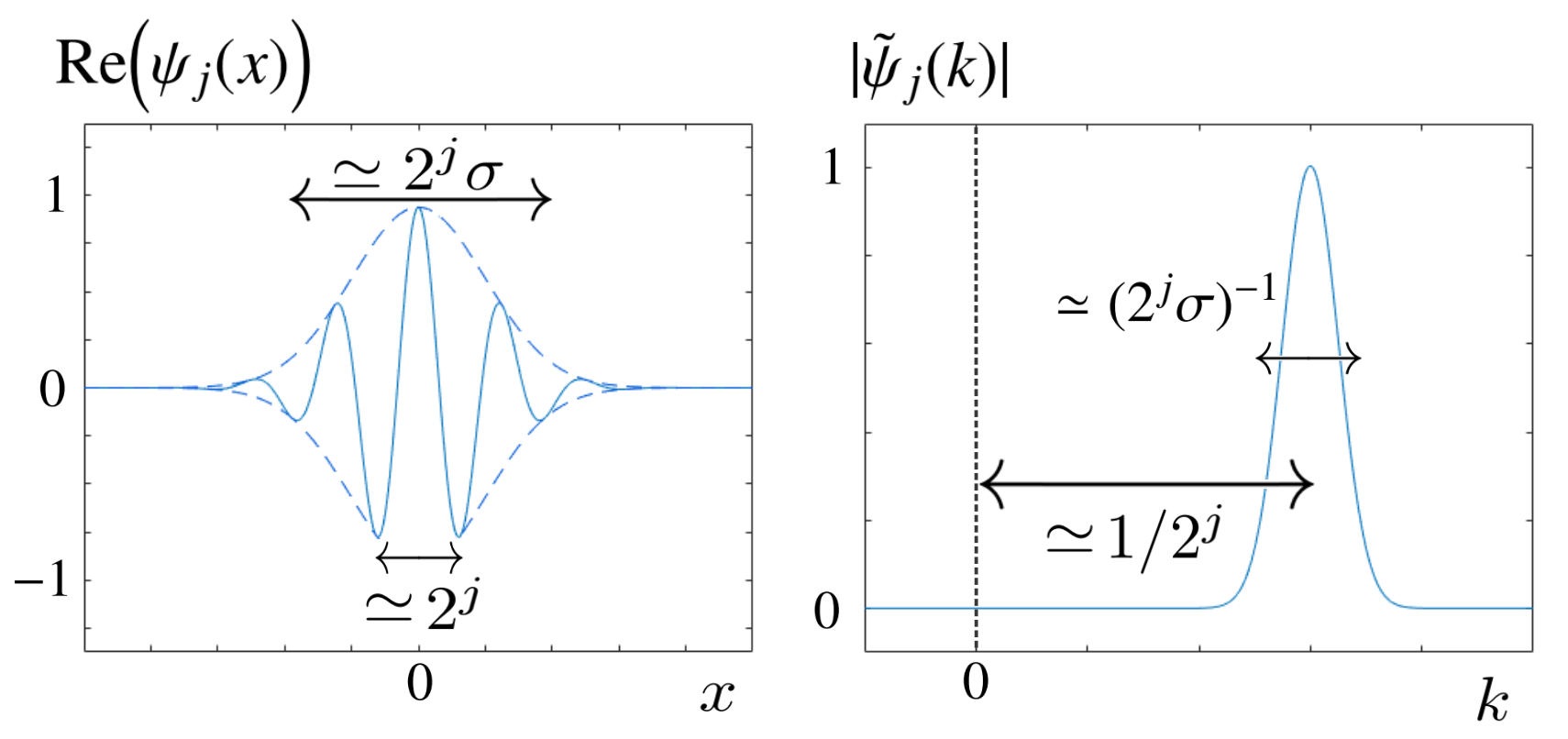}
\vspace{-0.5cm}
\caption{Real part of a Morlet wavelet in one dimension $\psi_j(x)$ (left), and amplitude of its Fourier transform $\tilde{\psi}_j (k)$ (right), with $\sigma=5$.}
\label{FigWavelet1D}
\end{center}
\end{figure}

Morlet wavelets in two dimensions can also be constructed from a mother wavelet, which is then dilated and rotated, as expressed in Eq.~\eqref{EqWavelets2D}. In this case, the mother wavelet is the generalization of the one-dimensional definition\footnote{In practice, the envelope of the oscillation is an elliptical Gaussian window to increase its angular resolution~\citep{scatnetdoc}, but this does not fundamentally modify our discussion.}:
\begin{equation}
\label{EqMotherWavelet2D}
\psi(\boldsymbol{x}) = \alpha \left( e^{i \boldsymbol{n}\cdot \boldsymbol{x}} - \beta \right) \cdot e^{- |\boldsymbol{x}|^2/(2 \sigma^2)},
\end{equation}
where $\boldsymbol{n}$ is a unit vector defining the oscillation direction of the mother wavelet\footnote{This direction may for instance be the $x$ direction in a $(x,y)$ plane.}. The Fourier transform of such a wavelet is still close to a Gaussian, whose central position and width are modified by rotations and scalings. Two examples of such wavelets and the supports of their Fourier transforms in the Fourier plane are shown in Fig.~\ref{FigWavelet2D}. We note that we consider a discrete set of wavelets in this paper, since they are built from an integer number of rotations and scalings, labelled with the $j$ and $\theta$ indices introduced in Sec.~\ref{PartComputationWSTCoeff}.

The $\sigma$ parameter describes approximately the number of oscillations of the wavelet within its support, and allows a trade-off between their spatial and frequency resolutions. Indeed, small values of $\sigma$ allow to detect the modulation of a given wavelength at a scale close to the wavelength itself, but at the cost of a poorer frequency localization. When studying fields linked to astrophysical observations, as in this paper, we use small values of $\sigma$ ($\sigma=0.8$ in the present case). Indeed, when studying the structure of a filament in a direction perpendicular to it, the main modulation it contains defines the width of the filament itself, and contains only one oscillation. Conversely, when studying audio signals, it is more suitable to use wavelets with a large value of $\sigma$, since the modulation timescales are often large in comparison to the period of audible sounds. 

We use in this paper the Morlet wavelet as a main tool. All the calculations are however very close to what one could obtain with the Discrete Windowed Fourier Transform (DWFT). Indeed, the one-dimensional DWFT of wavevector $k$ of a field $I(x)$ is
\begin{equation}
S[I](k,x) = \int \dd y ~ I(y) g(y-x) e^{-i k y}, 
\end{equation}
with $g$ a normalised window function. Choosing for this window a Gaussian with appropriate width (see Eq.~\eqref{EqDefMorlet1D}), the DWFT reduces (up to a global phase and in the limit where $\beta$ is negligible) to the convolution with a wavelet $\psi_j$ such that $k=2^{-j}$. It is thus possible to compute the power spectrum in the range of frequencies of a given $\psi_j$ Morlet wavelet as
\begin{equation}
\label{EqDefPS}
PS[I]\left(k=2^{-j}\right) = \frac{1}{L} \int {| I \star \psi_j |}^2 \dd x,
\end{equation}
where $L$ is the size of the integration domain~\citep{mallat2012group}. This result, that can be generalised in two dimensions, emphasises the difference between the usual power spectrum and the scattering coefficients, the latter being computed with the $L_1$ norm (see Eq.~\eqref{EqCoeffm1}).

\section{Flows studied}
\label{AppendixFlowsStudied}
In this Appendix, we present in detail the three different types of fields that we have applied our analysis to.
 
\subsection{Fractional Brownian motions}

\begin{figure}[t]
\begin{center}
\includegraphics[width = 0.495\textwidth]{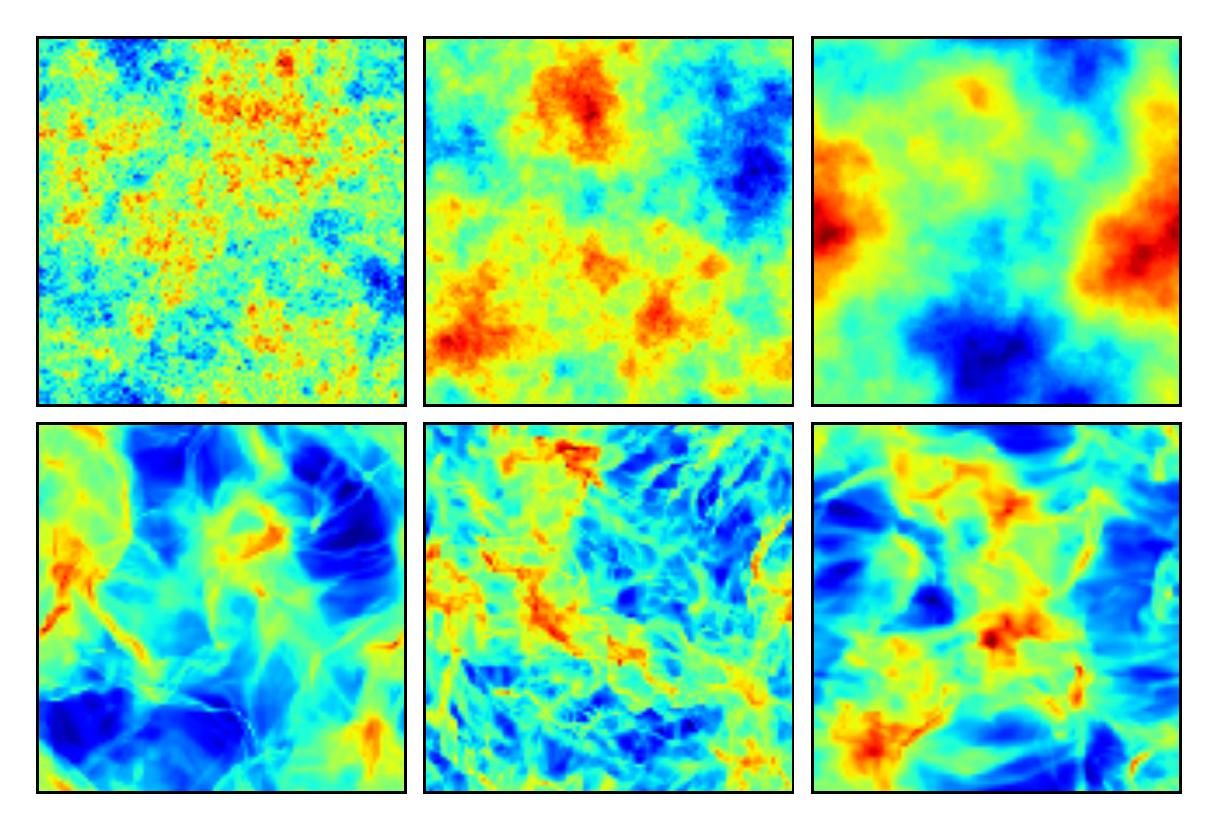}
\end{center}
\vspace{-0.1cm}
 \caption{\textit{Top}: 256$\times$256 maps of fBm processes, with $H=0.1$, 0.5, and 0.9 (from left to right). \textit{Bottom}: 256$\times$256 column density maps from snapshots of various MHD simulations (classes 1, 3, and 7 from left to right).}
\label{FigMapsfBmTurbFull}
\end{figure}

The two-dimensional purely synthetic random fields that we use in this work are fractional Brownian motions~\citep{falconer2004fractal}. These extend the class of Brownian motion processes by relaxing the condition of independent increments. In other words, values of fBm fields at nearby points are not independent, and this process is continuous but almost nowhere differentiable. In one dimension, an fBm of Hurst exponent $H\in\left]0,1\right[$  is defined as a random process $X:\mathbb{R}_+\mapsto\mathbb{R}$ such that the increments $X(t+\delta)-X(t)$ for any $t\geqslant 0$ and $\delta>0$ are normally distributed with zero mean and variance $\delta^{2H}$. In $N$ dimensions, a random field $X$ defined on $\mathbb{R}^N$ is an fBm if $\langle\left[X\left(\boldsymbol{r}_2\right)-X\left(\boldsymbol{r}_1\right)\right]^2\rangle\propto{||}\boldsymbol{r}_2-\boldsymbol{r}_1{||}^{2H}$, for any pair of points $(\boldsymbol{r}_1,\boldsymbol{r}_2)$. 

The syntheses of such fields are most easily built in Fourier space, with $\widetilde{X}(\boldsymbol{k})=A(\boldsymbol{k})\exp{\left[i\phi_X(\boldsymbol{k})\right]}$, 
by specifying amplitudes that scale as a power-law of the wavenumber $k=||\boldsymbol{k}||$, i.e. $A(\boldsymbol{k})=A_0k^{-\beta_X/2}$, where $\beta_X=2H+N$ is the spectral index. The Fourier phases $\phi_X$ are drawn from a uniform random distribution in $[-\pi,\pi]$, subject to the constraint $\phi_X(-\boldsymbol{k})=-\phi_X(\boldsymbol{k})$ so that $X$ is real-valued. The power spectra of fractional Brownian motions are therefore power laws, $P_X(k)\propto k^{-\beta_X}$. Three examples of such fields are given in Fig.~\ref{FigMapsfBmTurbFull} (\textit{top row}), with Hurst exponents equal to 0.1, 0.5 and 0.9.

In an astrophysical context, fBms have been used previously as toy models for the fractal structure of molecular clouds, in both density and velocity space~\citep{stutzki1998fractal,brunt2002interstellar,miville2003use}. They have also recently been used to model the turbulent component of the interstellar magnetic field, and to study the statistical properties of polarised thermal dust emission maps~\citep{levrier-et-al-2018}.

\subsection{Isothermal MHD simulations}

The second class of fields used in this work are column density maps $N_\mathrm{H}$ computed from numerical simulations of magnetohydrodynamical turbulent flows, aiming at reproducing the structures emerging in the interstellar medium. These simulations are performed by solving numerically the equations of ideal MHD, as described in \cite{Iffrig17}.

The simulations used in the present paper are simplified and do not take self-gravity into account. Stellar feedback (from supernovae and {\sc Hii} regions) is removed accordingly. Because the equations are solved on a finite-resolution grid, numerical diffusion mimics the effects of physical viscosity and dissipates energy in the fluid. Due to this dissipation, the simulations require constant energy input to attain a statistical steady state. In~\cite{Iffrig17}, the energy was injected by the stellar feedback, but here 
this is done through a turbulent forcing of the velocity field similar to the one described in~\cite{Schmidt09}. This forcing is quantified by the overall turbulent velocity dispersion $\sigma_\mathrm{turb}$. The thermodynamical treatment of the gas is simplified by assuming isothermality. Initially, the simulation cube is filled by a uniform-density, uniform-temperature gas, with $n_\mathrm{H}=2\,\mathrm{cm^{-3}}$ and $T=10\,\mathrm{K}$, and is permeated by a uniform magnetic field $B_0$. 

Several simulations are run with varying intensities of the magnetic field\footnote{Note that the values given here are smaller than the usually assumed values in the ISM, which are closer to 5\,$\mu\mathrm{G}$.}, from $B_0=0$ (hydrodynamical case) 
to $B_0=1\,\mu\mathrm{G}$, 
and of the turbulent forcing, with $\sigma_\mathrm{turb}=1\,\mathrm{km \, s^{-1}}$, to $\sigma_\mathrm{turb}=9\,\mathrm{km\,s^{-1}}$. To distinguish the different simulations, we group them into classes, as indicated in Table~\ref{Tab:MHDclasses}.

\begin{table}
\caption{Classes of MHD simulations}
\label{Tab:MHDclasses}                            
\centering
\begin{tabular}{c c c }
\hline\hline
Class & $B_0$ [$\mu$G] & $\sigma_\mathrm{turb}$ [$\mathrm{km\,s^{-1}}$]  \\    
\hline
1 & 0.0 & 1.0\\
2 & 0.0 & 4.0\\
3 & 0.0 & 9.0\\
4 & 0.5 & 1.0\\
5 & 0.5 & 4.0\\
6 & 0.5 & 9.0\\
7 & 1.0 & 1.0\\
8 & 1.0 & 4.0\\
9 & 1.0 & 9.0\\
\hline
\end{tabular}
\end{table}

Snapshots of the logarithm of total column density ($\log{N_\mathrm{H}}$) for several of these simulations are shown in Fig.~\ref{FigMapsfBmTurbFull} (\textit{bottom row}). The degree of anisotropy in these maps increases with the ratio between the mean value of the magnetic field and the turbulent forcing. We thus expect maps derived from simulations in class 3 to be isotropic, while those derived from simulations in classes 4 and 7 present higher levels of anisotropy.

\subsection{Herschel Polaris observations}
\label{AppendixPolaris}

\begin{figure*}[t]
  	\centering
  	\includegraphics[width = 0.99\textwidth]{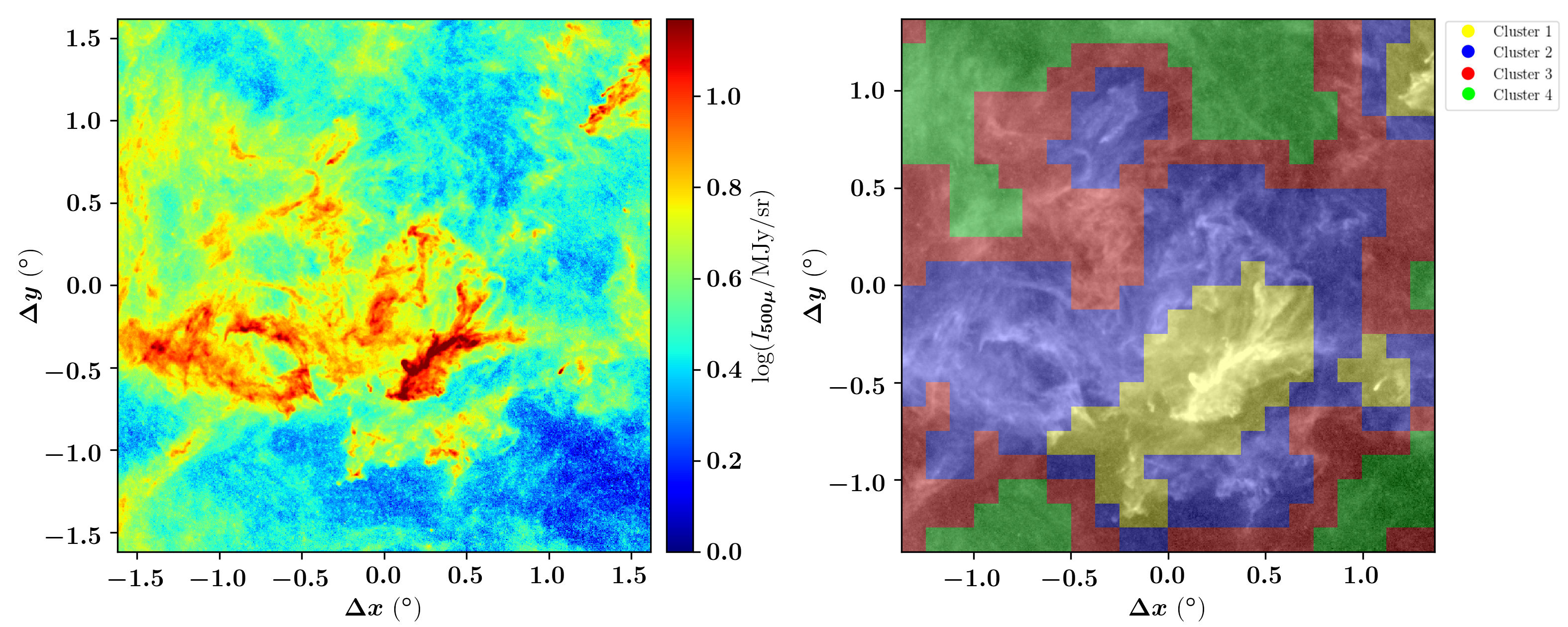}
	\caption{\textit{Left}: Dust thermal emission in the PolarisFlare observed with \textit{Herschel}-SPIRE-LW at $500\,\mu\mathrm{m}$~\citep{miville2010herschel}. The initial observation has been reshaped and the orientation of the axes is arbitrary. \textit{Right}: $k$-means clustering of the {\it Herschel} map in WST space, with $k = 4$. A cluster (green, red, blue, yellow colours) gathers regions of the map that have similar WST coefficients.}
	\label{FigPolarisFull}
\end{figure*}

The third field on which we have tested our approach is an observation of the Polaris Flare molecular cloud~\citep{miville2010herschel} obtained with the SPIRE instrument~\citep{griffin2010herschel} onboard the {\it Herschel} satellite~\citep{pilbratt2010herschel}, at a wavelength of 500\,$\mu\mathrm{m}$. The Polaris Flare is a diffuse molecular cloud that is not showing clear signs of star-formation activity. As such, it is generally thought to be representative of the very early stages of molecular cloud formation and evolution, and the dynamics of its gas and dust contents are therefore probably more representative of the interstellar turbulent cascade than other, star-forming clouds, in which feedback processes from young stars (jets, outflows, and radiation) tend to confuse the picture. 

The far-infrared emission that was mapped by {\it Herschel}-SPIRE at a resolution of 37$\arcsec$, is produced by the cold dust in the cloud, as it reprocesses the ambient visible and UV radiation from Galactic starlight. At these long wavelengths, this emission is optically thin, and its integrated intensity is therefore directly proportional (to a very good level of approximation) to the column density of the large, cold grains on the line of sight. It allows to probe the matter content of the cloud, assuming a uniform gas-to-dust ratio. We note that the Polaris Flare has been extensively studied, not only through this thermal continuum emission of cold dust, but also through CO rotational lines that allow to probe the velocity field of the molecular gas down to very small scales~\citep{falgarone1998iram,hily2009intermittency}. The geometry of the magnetic field in the Polaris Flare was also studied with optical stellar polarisation data by~\cite{panopoulou2016magnetic}. 

We use a $832\times832$ pixels subset of the full {\it Herschel}-SPIRE map discussed in~\cite{miville2010herschel}, covering almost 10 square degrees in the sky (Fig.~\ref{FigPolarisFull}, left). Compared to the fBm and MHD simulations, the statistical properties of this map are unlikely to be homogeneous\footnote{For example, the filamentary structures just south of the centre of the map might be gravitationally bound, but the diffuse filaments towards the edge probably are not.}. It is therefore necessary to work with {\it local} WST coefficients and, ideally, to identify a mesoscopic scale over which the statistical properties may be considered homogeneous and study their variations over larger scales, as discussed in Appendix~\ref{AppendixLocalWST}.

To circumvent this difficulty, we propose, as a first attempt to distinguish between a spatial evolution of the statistical properties across the Polaris Flare and the statistical variability that is intrinsic to an homogeneous stochastic process, to compute local WST coefficients on the map (using a Gaussian window $\phi_J$ of width $2^J=64$ pixels for $J=6$, see Appendix~\ref{AppendixLocalWST}), and gather regions in the sky that have similar WST coefficients using a clustering algorithm\footnote{Note that the local WST coefficients are computed with some oversampling, which means that the windows on which they are computed partially overlap~\citep{scatnetdoc}. Due to these local windows, it is also necessary to exclude a thin band close to the edges of the map.}. To do so, we divide the Polaris Flare map into $N = 22\times22$ square regions and for each region we compute a set of normalised WST coefficients that we note $\mathbf{y_i}$. These $\mathbf{y_i}$ can be seen as a vector in a statistical space of dimension $1009$ (with $J = 6$ and $L = 8$). To identify regions that have similar WST coefficients, we use a $k$-means clustering algorithm\footnote{Note that this clustering approach has already been used in studies of the interstellar medium~\citep{2018A&A...610A..12B}.} which performs a partition of $\{\mathbf{y_i}\}_{k\in\{1, ..., N\}}$ in $k$ subsets $\{S_1, ..., S_k\}$ so that the sum of the variances of Euclidean distances between vectors within each cluster is minimised~\citep{arthur2007k}. Formally, the $k$-means algorithm finds a partition which minimises:
$$
	\sum_{j=1}^k\sum_{\mathbf{y_i}\in S_j}||\mathbf{y_i}-\mathbf{\mu_j}||^2
$$
where $\mathbf{\mu_j}$ is the centroid of cluster $S_j$.

Figure~\ref{FigPolarisFull} (right) shows the four clusters of the Polaris Flare map that were identified in this way. This number of clusters is a free parameter of the algorithm, but we have not explored its influence, settling for $k=4$ as a first guess. It already shows a clear distinction between the statistical properties of the identified clusters. For each of these regions, we then assume that the statistical properties are homogeneous, so that local WST coefficients within each region can be averaged.  


\section{Possible additional terms to the RWST angular reduction}
\label{AppendixDetailsDuFit}

In Sec.~\ref{PartModelMain}, we discussed the form of the dominant terms accounting for the angular dependencies of the WST coefficients (Eqs.~\ref{EqScatDecomp1} and~\ref{EqScatDecomp2}), but it may happen that residuals exhibit oscillatory trends with different periodicities, showing that these terms are not sufficient. This is mostly apparent when averaging over several realizations, because these residual trends then start to stand out from the sampling noise. In this case, several additional terms are used to satisfactorily fit these minor features.

When fitting the angular dependency of WST coefficients averaged over twenty 256$\times$256 maps, we identified three such additional terms, that are either new angular modulations due to additional physical effects, or higher harmonics of an angular modulation that has already been identified. We stress, however, that these terms have small amplitudes compared to the RWST coefficients discussed in the main text, and that the values of the latter are unaffected by the inclusion of these additional terms in the fit. The discussion of these additional minor terms nevertheless offers a better understanding of the limits of the dominant terms discussed in the main text.

The first two additional terms are modulations related to pixelation. These terms are not attributable to the WST computation, but to the methods used to generate the fields, and may be different for the different types of fields. For instance, we may expect a signature at small scale of the grid that the MHD simulations were computed on. Similarly, the computation of the fBm fields on a square, regular grid in Fourier space should produce signatures at both small and large scales. Similar effects related to pixelation have been identified in the map of the Polaris Flare.

\begin{figure}[t]
\begin{center}
\includegraphics[width = 0.48\textwidth]{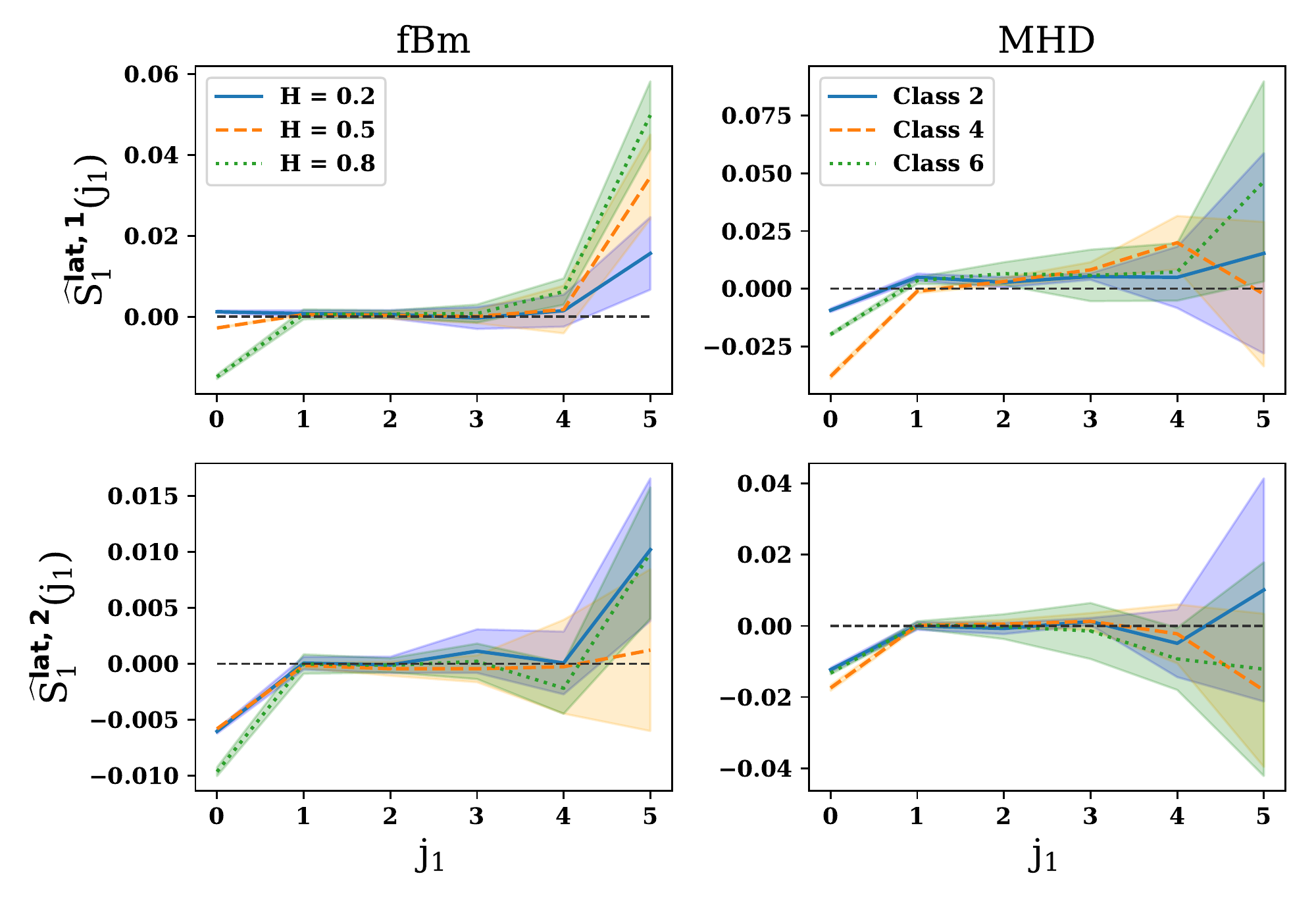}
\vspace{-0.5cm}
\end{center}
 \caption{Additional $ \hat{S}^\text{lat,1}_1 (j_1)$ and $ \hat{S}^\text{lat,2}_1 (j_1)$ terms related to lattice pixelation, for the $m=1$ layer. For each class of fBm processes or MHD simulations, these were computed using twenty $256\times256$ maps.}
\label{FigPlotSCR1_Res}
\end{figure}

Following the discussion in Sec.~\ref{PartModelMain}, we expect any modulation related to the lattice to be $\pi/2$-periodic and aligned with the lattice's main directions, that is with a reference angle $\theta_r=1$. Experience shows that the second harmonic also needs to be taken into account. For the $m=1$ and $m=2$ layers, we therefore respectively add the following terms to the decompositions given in Eq.~\eqref{EqScatDecomp1} 
\begin{multline}
\label{EqAddTerm1}
 \cdots ~  + ~ \hat{S}^\text{lat,1}_1 (j_1) \cdot \cos \Big( \frac{4 \pi}{\Theta} \big[ \theta_1 - 1]\Big) \\
+  ~ \hat{S}^\text{lat,2}_1 (j_1) \cdot \cos \Big( \frac{8 \pi}{\Theta} \big[ \theta_1 - 1]\Big),
\end{multline}
and in Eq.~\eqref{EqScatDecomp2}
\begin{multline}
\label{EqAddTerm2}
 \cdots ~  + ~ \hat{S}^\text{lat,1}_2 (j_1,j_2) \cdot \cos \Big( \frac{4 \pi}{\Theta} \big[ \theta_1 - 1]\Big) \\
+  ~ \hat{S}^\text{lat,2}_2 (j_1,j_2) \cdot \cos \Big( \frac{8 \pi}{\Theta} \big[ \theta_1 - 1]\Big).
\end{multline}

When included, these additional terms have low levels (see Fig.~\ref{FigPlotSCR1_Res} for examples), and all the fields but the fBm show such non-vanishing additional components only at the smallest scales. A signature of the lattice also seems to appear at large scales for fBm fields, but this is weakly significant and difficult to precisely assess. We note that the $\pi/2$ and $\pi/4$ harmonics have similar amplitudes. This is not surprising since anisotropic terms related to lattice pixelation may be much less smooth than a physical anisotropy. 

The third additional term that we have identified is a $\pi/2$ harmonic for the $ \hat{S}_2^\text{iso,2}$ component of Eq.~\eqref{EqScatDecomp2}, that needs to be added to this equation 
\begin{equation}
\label{EqAddTerm3}
 \cdots ~  + ~ \hat{S}^\text{iso,3}_2 (j_1,j_2) \cdot \cos \Big( \frac{4 \pi}{\Theta} \big[ \theta_1 - \theta_2]\Big).
\end{equation}
The appearance of such a term in the angular reduction of the WST coefficients is in line with the discussion of Sec.~\ref{PartModelMain} about the structure of the angular modulations. The $ \hat{S}^\text{iso,3}_2$ terms may also contain additional information on the fields. For instance, Fig.~\ref{FigRSC2_Iso2} shows that the fBm fields and MHD simulations clearly have different forms for this $\pi/2$ harmonics, providing yet another quantitative lever to compare various processes.

No other additional term was necessary to achieve satisfactory fits of the residual trends, but future studies may need to include further terms of a similar type (e.g. $\pi/2$ harmonics for the anisotropic $\hat{S}_1^\text{aniso}$, $\hat{S}_2^\text{aniso,1}$ and $\hat{S}_2^\text{aniso,2}$ terms). However, we expect all these additional terms to remain small compared to the main RWST coefficients described in Sec.~\ref{PartModelMain}.

\begin{figure}[t]
\begin{center}
\includegraphics[width = 0.50\textwidth]{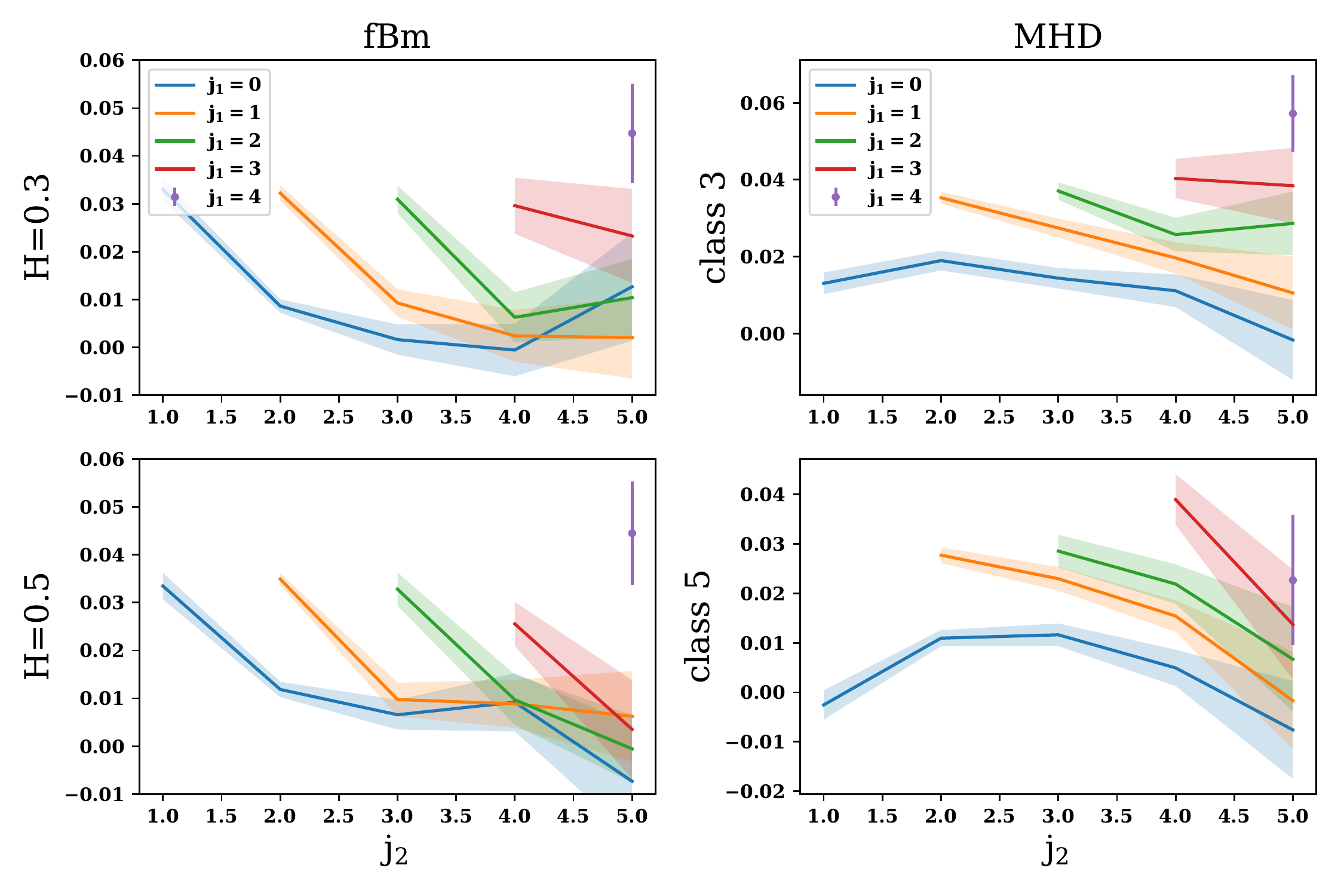}
\vspace{-0.5cm}
\end{center}
 \caption{Additional $ \hat{S}^\text{iso,3}_2 (j_1,j_2)$ terms. For each class of fBm processes or MHD simulations, these were computed using twenty $256\times256$ maps.}
\label{FigRSC2_Iso2}
\end{figure}

\section{Heterogeneous statistics and the local wavelet scattering transform}
\label{AppendixLocalWST}
In this Appendix, we discuss the generalisation of the WST to fields that are not statistically homogeneous and how it can be applied locally in such cases.

\subsection{Mesoscopic scale}
The use of any statistical measure to describe properties of fields arising from nonlinear physical processes, and hopefully, from there, to gather information about the underlying physics itself, warrants some discussion about the physical quantities that may be encoded in the statistics, and about the scales over which these statistics are computed. A useful analogy here can be found in statistical thermodynamics, which proceeds by devising statistical measures over a vast number of particles to establish physically meaningful quantities\footnote{For instance, the kinetic temperature arises as the parameter characterizing the dispersion of particle velocities in ideal gases.}. These averages are performed over {\it mesoscopic} scales, large enough to contain a huge number of particles so that statistical fluctuations may be neglected, but also small enough so that the thermodynamical variables can be seen as locally-defined fields. These may in turn vary over larger, macroscopic scales.

In our case, we have observable fields, such as column density maps, whose morphologies we mean to describe statistically, with the purpose of relating these statistics to physical properties of the medium, such as the amplitude of the magnetic field averaged on a certain scale. Assuming that such a relationship exists, it is important to ask which scales and structures in the observable fields are subject to the inherent variability that is meant to be captured by the statistics, and which ones are related to a modification of the larger scale physical properties associated in turn with the statistical properties themselves. Our mesoscopic scale should be chosen where these two ranges of scale meet, so that the statistics can be considered homogeneous below this scale while allowing for a subsequent study of the variation of physical properties across larger scales\footnote{An important difference between our case and statistical thermodynamics is that in the latter the quantities defined on a mesoscopic scale (e.g. kinetic temperature) are different in nature from those they are built upon at the microscopic scale (e.g. particle velocities), while in our case, these two quantities could well be the same, for example the amplitude of the magnetic field. With a statistical description that allows a scale separability, as it is the case for the WST and its reduced form, we can treat the variations of these quantities statistically at small scales, and relate these statistics to the value of the same physical quantities averaged on the mesoscopic scale.}. 

It may be difficult to properly determine this mesoscopic scale. A good criterion seems to be the apparent reproduction of similar patterns. For instance, in the Polaris Flare map (Fig.~\ref{FigPolarisFull}), structures at the scale of $30\arcmin$ are too scarce to be treated statistically, but those appearing at the $0.3\arcmin$ scale may be. The mesoscopic scale required to describe such a map should then be somewhere in between these two scales, which is why we chose $J=6$, corresponding to $15\arcmin$. We see that the scale separability provided by the WST and the RWST is of great importance.

\subsection{The local WST}
When the field considered has homogeneous statistical properties, it is sufficient to compute a set of global scattering coefficients that can be obtained by integration on the entire spatial support of this field (Eqs~\eqref{EqCoeffm0}-\eqref{EqCoeffm2}). It is however also possible to compute scattering coefficients that describe local statistical properties~\citep{bruna2013invariant} on mesoscopic scales. In the homogeneous case, these local WST coefficients provide different samples and allow us to estimate the variance. In the heterogeneous case, they allow us to quantify the evolution of statistical properties on large scales.

For a real-valued field $I(\boldsymbol{x})$, these local WST coefficients $S^\text{loc}_0 (\boldsymbol{x})$, $S^\text{loc}_1 [j_1,\theta_1] (\boldsymbol{x})$ and $S^\text{loc}_2 [j_1,\theta_1,j_2,\theta_2] (\boldsymbol{x})$ are computed similarly to the global coefficients but with a spatial integration limited to a subset of the space, using a normalised Gaussian window $\phi_J$ of fixed width $2^J$, the size of the largest wavelength probed by the Morlet wavelets~\citep{bruna2013invariant}. The $m=0$ coefficient is the local average of the field over a characteristic scale $2^J$, i.e.
\begin{equation}
S^{\text{loc}}_0 (\boldsymbol x)= \left[ I \star \phi_J \right] (\boldsymbol x),
\end{equation}
while the $m=1$ and $m=2$ coefficients are given by
\begin{equation}
S^{\text{loc}}_1 [j_1,\theta_1] (\boldsymbol x) = \mu_{1,\text{loc}}^{-1} \big[ | I \star \psi_{j_1,\theta_1} | \star \phi_J \big] (\boldsymbol x),
\end{equation}
and
\begin{equation}
S^{\text{loc}}_2 [j_1,\theta_1,j_2,\theta_2] (\boldsymbol x) = \mu_{2,\text{loc}}^{-1} \big[ || I \star \psi_{j_1,\theta_1} |  \star \psi_{j_2,\theta_2} |\star \phi_J  \big] (\boldsymbol x),
\end{equation}
where the $\mu_{i,\text{loc}}$ normalization factors are the $m=1$ and $m=2$ responses to a Dirac $\delta$ function,
\begin{equation}
\mu_{1,\text{loc}}=  \big[ | \delta \star \psi_{j_1,\theta_1} | \star \phi_J \big] (\boldsymbol x),
\end{equation}
and similarly for $ \mu_{2,\text{loc}}$. The normalization described in Sec.~\ref{PartWSTNumberNorm} [Eqs.~\eqref{EqNormalize1} and~\eqref{EqNormalize2}] can be performed at this stage. Then, the computation of local RWST coefficients can be done following exactly the computation described in Sec.~\ref{PartReducedScatCoeff}. Otherwise, integrating these local coefficients over the entire space recovers the global scattering coefficients. 

\section{Additional results}
\label{AppendixCompleteSets}

We give in this appendix additional sets of RWST coefficients for the various processes studied in this paper, as well as supplementary examples of RWST syntheses. The reduced scattering coefficients given in Figs.~\ref{FigS1_Full} to~\ref{FigS2_Full_Pol} have been obtained from sets of twenty $256\times256$ maps of MHD simulations or fBm processes, as well as from the four clusters in the Polaris Flare map (see Appendix~\ref{AppendixFlowsStudied}). The coefficients given in Figs.~\ref{FigS2_Full_fBm} to~\ref{FigS2_Full_Pol} use the angular fits given by Eqs.~(\ref{EqScatDecomp1}) and~(\ref{EqScatDecomp2}), while the ones given in Fig.~\ref{FigS1_Full} also include the additional terms described in Appendix~\ref{AppendixDetailsDuFit} [Eqs.~\eqref{EqAddTerm1} to~\eqref{EqAddTerm3}]. 

In Fig.~\ref{FigSynthFull}, we show additional syntheses. These are based on the RWST coefficients derived from the WST coefficients averaged over twenty maps for the MHD and fBm processes, and over each cluster for the Polaris Flare map. They are produced following the method described in Sec.~\ref{PartSecSynthesis}. We display the synthetic fields obtained from the RWST coefficients of the different Polaris Flare clusters next to 256$\times$256 zooms of the original 832$\times$832 Polaris Flare map, covering regions where the given cluster is dominant. The regions in that zoom that do not belong to this specific cluster are shaded. We note that, especially because these fields are not statistically homogeneous, they present a much larger dynamic range in terms of local averages. To allow satisfactory visual comparisons, we therefore subtracted the mean value of the 256$\times$256 maps we show\footnote{This was applied both to the original maps and to the synthetic ones. The same colour scale is used in both.}. If the synthesis of cluster 4 is satisfactory (except at the largest scales, as already discussed), that for cluster 2 shows the limitation of the clustering performed, because the statistical properties of the field seem to vary with the local mean level. 


\bibliographystyle{aa}

\begin{figure*}[t!]
\begin{center}
\includegraphics[width = 0.99\textwidth]{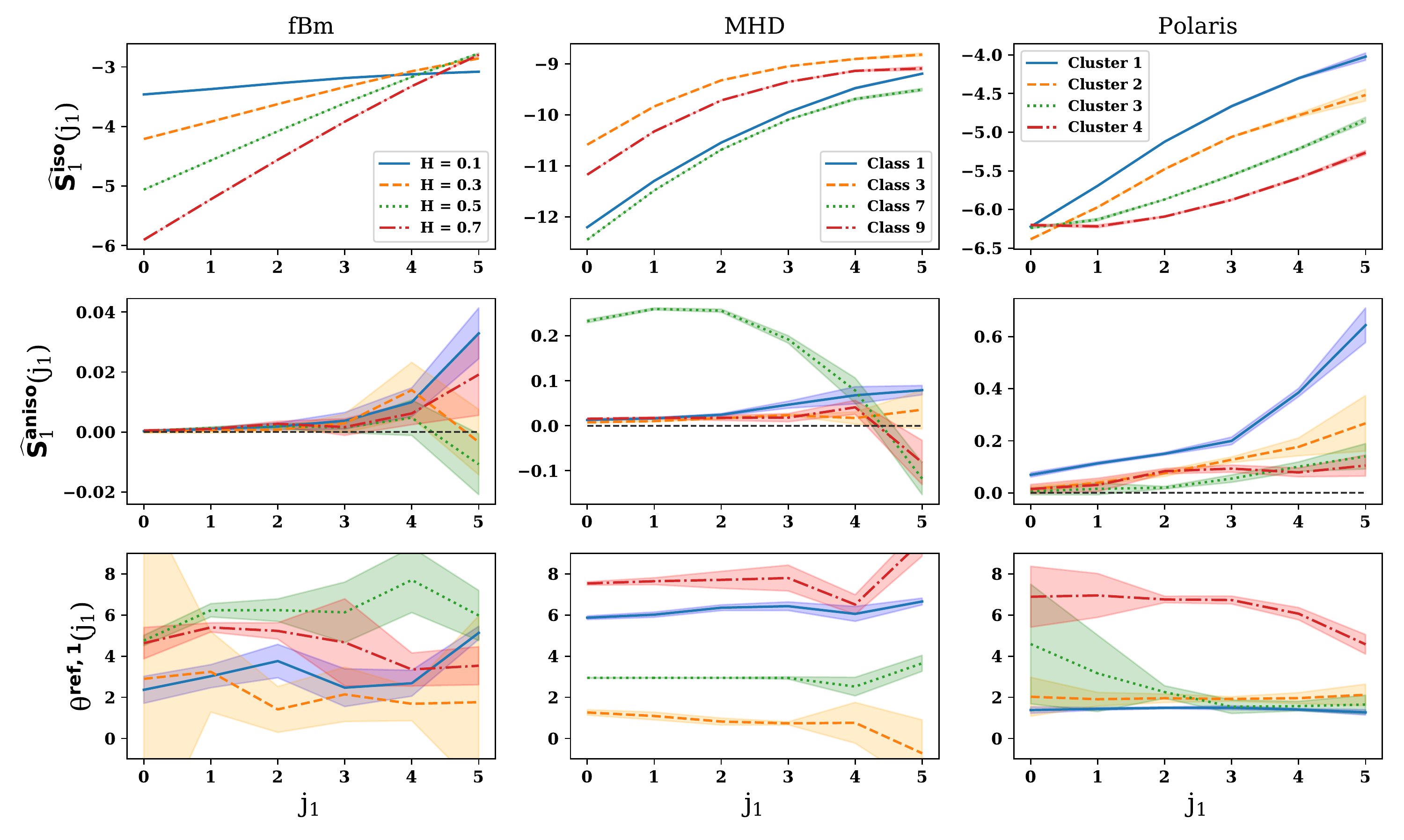}
 \caption{Plots of $m=1$ reduced scattering coefficients for four class of synthetic fBm processes, MHD simulations, and the four clusters of the \textit{Herschel} Polaris observation. These coefficients have been obtained from sets of twenty $256\times256$ maps, using the reduction given in Sec.~\ref{PartModelMain} as well as the additional reduced terms discussed in Appendix~\ref{AppendixDetailsDuFit}.}
 \label{FigS1_Full}
 \end{center}
\end{figure*}

\begin{figure*}[t!]
\begin{center}
\includegraphics[width = 0.99\textwidth]{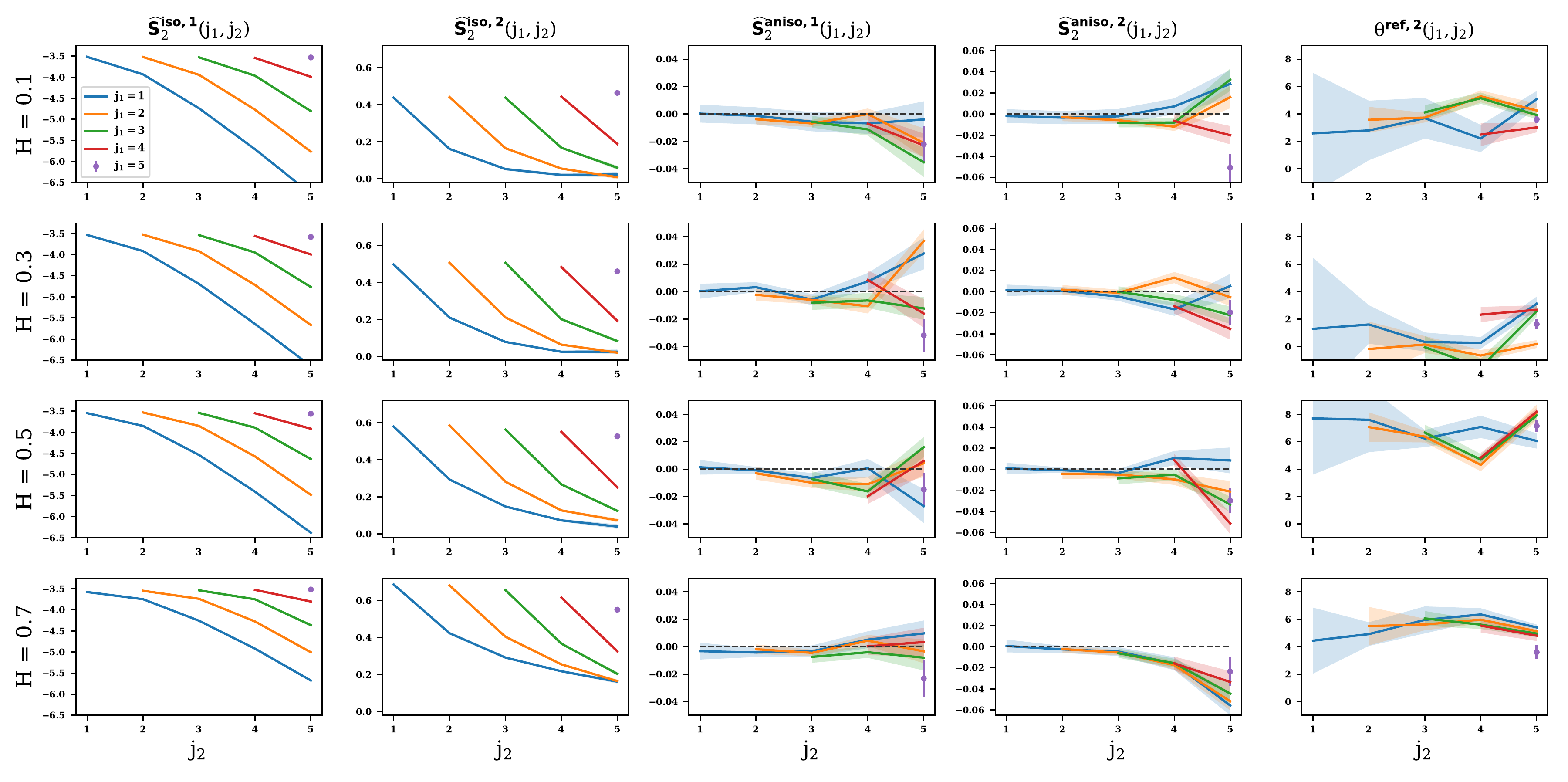}
 \caption{RWST coefficients for $m=2$ and four classes ($H=0.1$, $H=0.3$, $H=0.5$, $H=0.7$) of fBm processes. For each class, twenty $256\times256$ maps are used. Each curve correspond to a fixed $j_1$ value, and $j_2$ values from $j_1+1$ to 5.} 
 \label{FigS2_Full_fBm}
 \end{center}
\end{figure*}

\begin{figure*}[t!]
\begin{center}
\includegraphics[width = 0.99\textwidth]{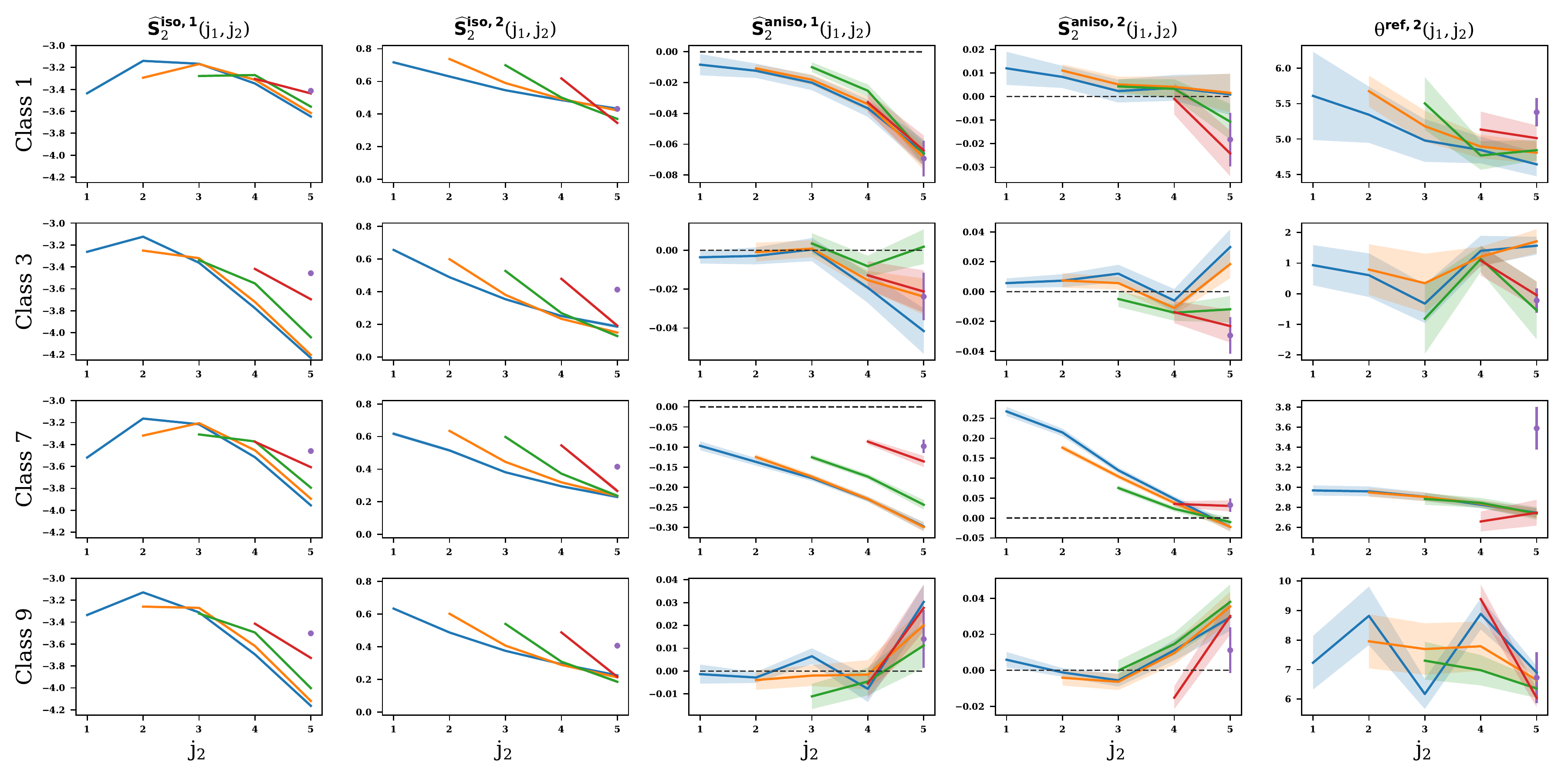}
 \caption{Same as Fig.~\ref{FigS2_Full_fBm} but for four classes (1, 3, 5, 7) of MHD simulations. For each class, twenty $256\times256$ maps are used.}
 \label{FigS2_Full_Turb}
 \end{center}
\end{figure*}

\begin{figure*}[t!]
\begin{center}
\includegraphics[width =0.99\textwidth]{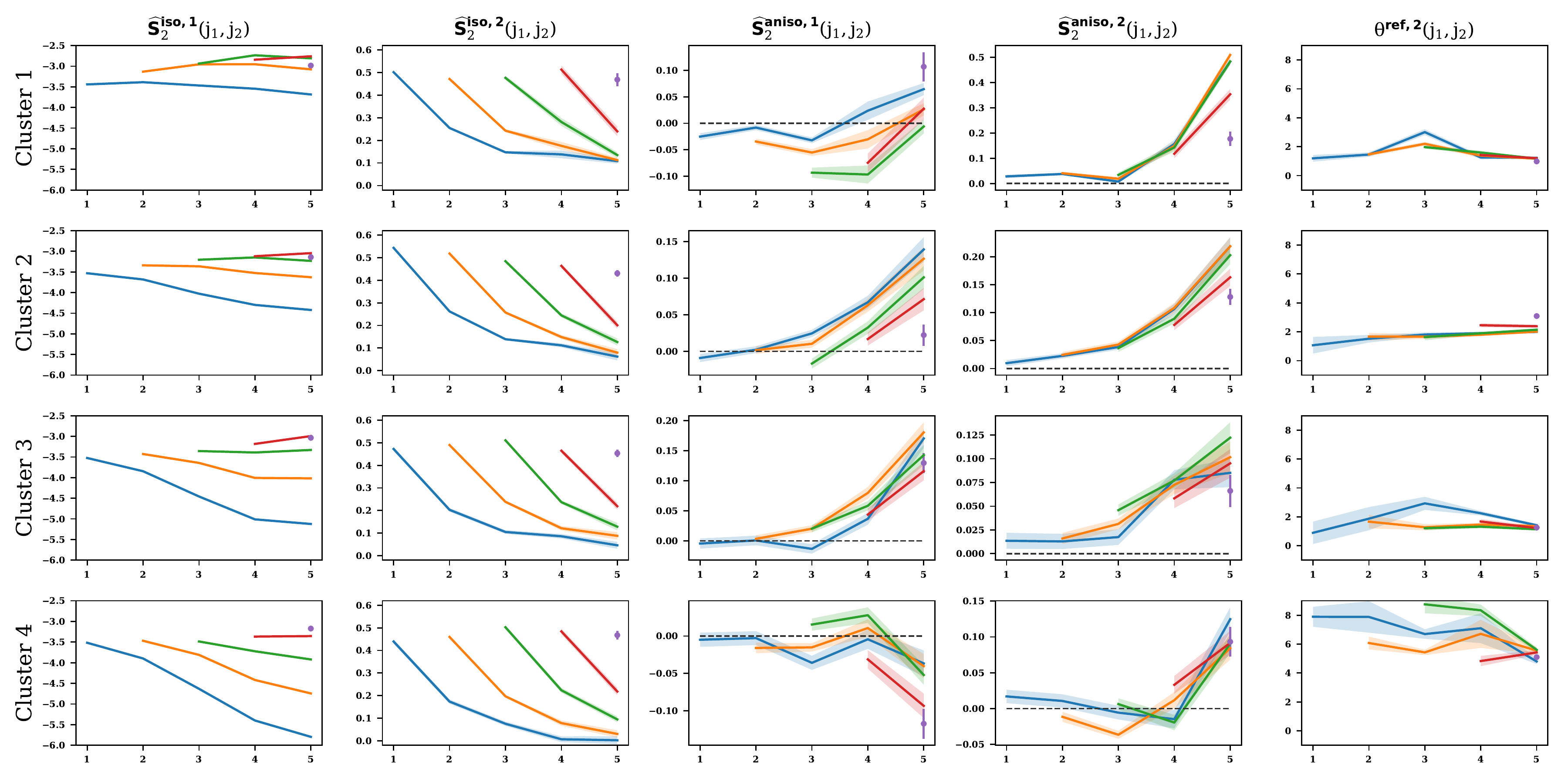}
 \caption{Same as Fig.~\ref{FigS2_Full_fBm} but for the four clusters of the Polaris Flare map.}
 \label{FigS2_Full_Pol}
 \end{center}
\end{figure*}

\begin{figure*}[t!]
\begin{center}
\includegraphics[width =0.92\textwidth]{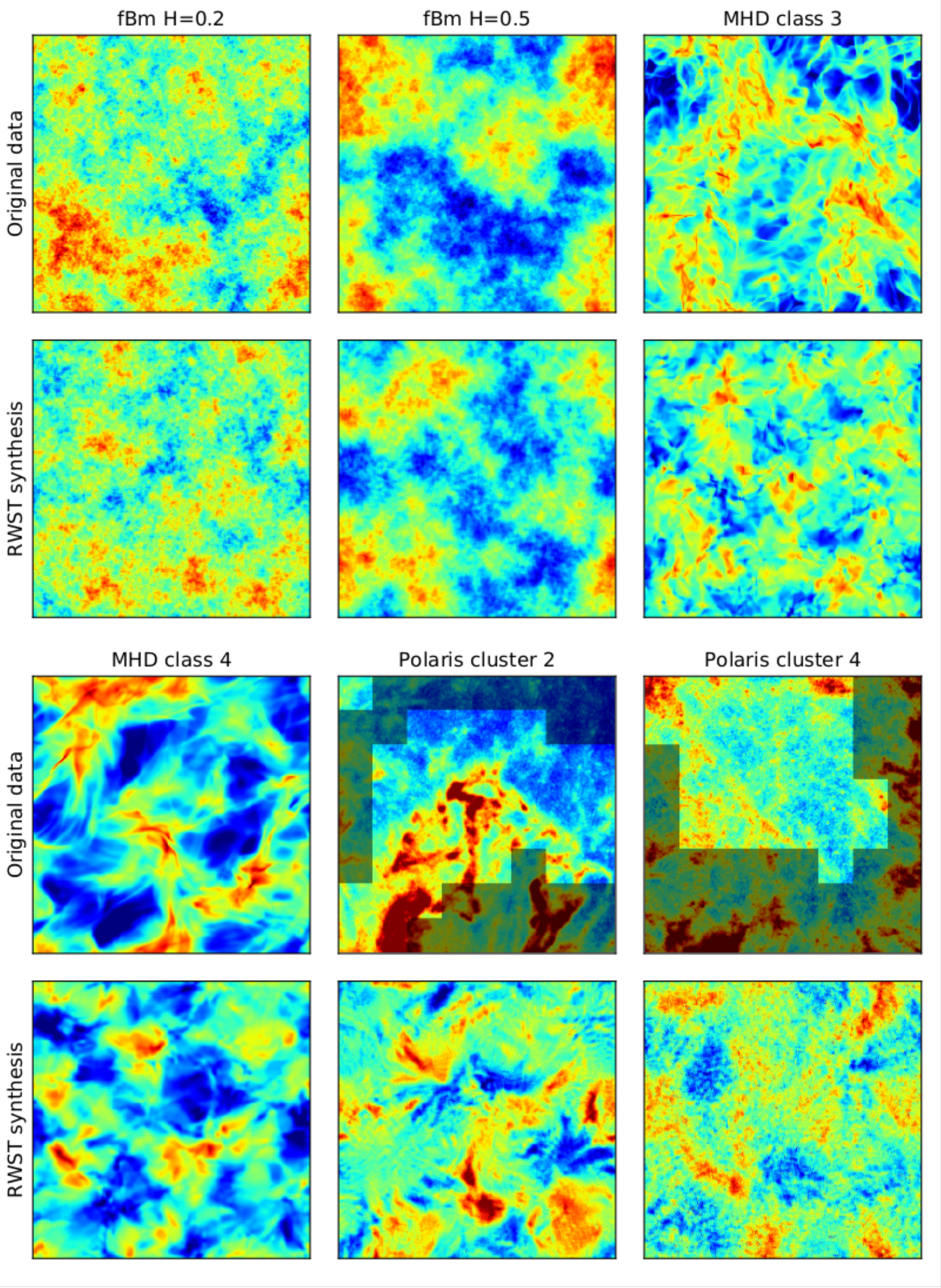}
 \caption{Additional examples of RWST syntheses. The RWST coefficients have been obtained from twenty 256$\times$256 maps for the fBm synthetic fields and the MHD simulations, and from the different clusters of the Polaris Flare map. The original data shown here for the Polaris Flare are 256$\times$256 subsets of the original map, and all regions but the cluster under study are shaded. The colourscale is the same for each pair of original and synthetic fields. For the Polaris Flare case, all maps are mean-subtracted.}
 \label{FigSynthFull}
 \end{center}
\end{figure*}

\end{document}